\documentclass[final,3p,times,twocolumn]{elsarticle}

\usepackage{graphicx}      
\usepackage{amsfonts}
\usepackage{gensymb}
\usepackage{amsmath}
\usepackage{amssymb}
\usepackage{amsthm}
\usepackage{algorithm}
\usepackage{algorithmic}
\usepackage{mathtools}
\usepackage{svg}
\usepackage[shortlabels]{enumitem}
\usepackage{soul} 
\usepackage{chemformula}
\usepackage[version=4]{mhchem}
\usepackage[makeroom]{cancel}
\usepackage{graphicx}   
\usepackage{float}      
\usepackage{subcaption} 
\usepackage{siunitx}
\usepackage[linktocpage=true, colorlinks=true, linkcolor=blue,citecolor=blue, pdftitle={EIS - review}, pdfauthor={Noel Hallemans}]{hyperref} 

\setstcolor{red}\setulcolor{red} 

\newcommand{\Hexc}{\mathbb{H}_\text{exc}}

\usepackage{nomencl}
\makenomenclature
\usepackage{etoolbox} 
\renewcommand\nomgroup[1]{%
  \item[\bfseries
  \ifstrequal{#1}{A}{Acronyms}{%
  \ifstrequal{#1}{C}{Constants}{%
  \ifstrequal{#1}{O}{Operators}{%
  \ifstrequal{#1}{S}{Sets}{%
  \ifstrequal{#1}{V}{Variables}{%
  \ifstrequal{#1}{F}{Functions}{}}}}}}%
]}

\journal{Electrochimica Acta}

\begin{document}

\begin{frontmatter}



\title{Electrochemical impedance spectroscopy beyond linearity and stationarity --- \\ a critical review
}


\author[inst1,inst2]{Noël Hallemans\corref{cor1}}

\cortext[cor1]{Corresponding author}
\ead{noel.hallemans@vub.be}
\author[inst7,inst8]{David Howey}
\author[inst3]{Alberto Battistel} 
\author[inst2]{Nessa Fereshteh Saniee} 
\author[inst4]{Federico Scarpioni}
\author[inst5]{Benny Wouters} 
\author[inst4,inst6]{Fabio La Mantia}
\author[inst5]{Annick Hubin}
\author[inst2,inst8]{Widanalage Dhammika Widanage}
\author[inst1]{John Lataire}

\address[inst1]{Research Group Fundamental Electricity and Instrumentation, Vrije Universiteit Brussel, Pleinlaan 2, 1050 Brussels, Belgium}
\address[inst2]{WMG, University of Warwick, Coventry, 7AL CV4, UK}
\address[inst7]{Battery Intelligence Lab, Department of Engineering, University of Oxford, OX1 3PJ, UK}
\address[inst8]{The Faraday Institution, Quad One, Harwell Science and Innovation Campus, Didcot, UK}
\address[inst3]{Institute of Technical Medicine, Furtwangen University, Jakob-Kienzle-Strasse 17, Villingen-Schwenningen 78054, Germany}
\address[inst4]{Fraunhofer Institute for Manufacturing Technology and Advanced Materials IFAM, Wiener Strasse 12, Bremen 28359, Germany}
\address[inst5]{Research Group Electrochemical and Surface Engineering, Vrije Universiteit Brussel, Pleinlaan 2, 1050 Brussels, Belgium}
\address[inst6]{Energy Storage and Energy Conversion Systems, Bremen University, Wiener Strasse 12, Bremen 28359, Germany}

\begin{abstract}
Electrochemical impedance spectroscopy (EIS) is a widely used experimental technique for characterising materials and electrode reactions by observing their frequency-dependent impedance. Classical EIS measurements require the electrochemical process to behave as a linear time-invariant system. However, electrochemical processes do not naturally satisfy this assumption: the relation between voltage and current is inherently nonlinear and evolves over time. Examples include the corrosion of metal substrates and the cycling of Li-ion batteries. As such, classical EIS only offers models linearised at specific operating points. During the last decade, solutions were developed for estimating nonlinear and time-varying impedances, contributing to more general models. In this paper, we review the concept of impedance beyond linearity and stationarity, and detail different methods to estimate this from measured current and voltage data, with emphasis on frequency domain approaches using multisine excitation. In addition to a mathematical discussion, we measure and provide examples demonstrating impedance estimation for a Li-ion battery, beyond linearity and stationarity, both while resting and while charging. 
\end{abstract}


\end{frontmatter}

Electrochemistry studies processes at electrode/electrolyte interfaces. These processes involve the movement of charged species (ions or electrons), generating a current flowing through a cell and a voltage drop over its electrodes. Diverse noninvasive techniques relying on current and voltage measurements have been developed for studying these processes. Typically, one of the two quantities is kept constant, swept, or oscillated, while the other quantity's response is recorded. Some of the most widely used techniques are linear sweep voltammetry, constant current chrono-potentiometry, constant-potential chronoamperometry and electrochemical impedance spectroscopy (EIS)\nomenclature[A]{EIS}{Electrochemical impedance spectroscopy}. In EIS \cite{EISbook,MacDonaldImpedanceSpectroscopy,gabrielli1984identification,gabrielli2020once,wang2021electrochemical}, the dynamics of electrochemical processes are studied by means of the impedance response to the applied current or voltage, and the term `spectroscopy'  refers to the frequency dependency. It is worth emphasising that EIS is a nonparametric data-driven technique, i.e.\ the impedance is solely computed relying on current and voltage data, without prior knowledge about the governing equations as in physics-based modelling \cite{doyle1993modeling,chen2020development,planella2021systematic}.

During the last two decades, EIS has gained wide popularity thanks to its accessible implementation and broad applicability to, among others, corrosion \cite{mansfeld1990electrochemical,mansfeld1995use,revilla2020eis}, batteries \cite{troltzsch2006characterizing,andre2011characterization,waag2013experimental,richardson2014battery,pastor2017comparison,meddings2020application,zhu2018electrochemical,zhang2020identifying,gabervsvcek2021understanding,jones2022impedance}, and fuel cells \cite{he2009exploring,niya2013study}. Nowadays, EIS is available in many commercial cyclers and potentiostats, where a user decides on a set of frequencies and the device measures the complex impedance values at these frequencies.

From a system theoretical perspective, the impedance is a special case of a transfer function, which is a model for a linear time-invariant (LTI)\nomenclature[A]{LTI}{Linear time-invariant} dynamical system \cite{kailath1980linear}. In system theory, dynamical systems are systems with memory, that is, systems defined through differential equations or, equivalently, convolution operators. This is opposed to static systems, where the output is simply a static function of the input. This interpretation of the term dynamic should not be confused with its use in electrochemistry to denote time-variation. In this article, we use the system theory convention. Models relate the output of the system to its input, which, in the EIS case, is the current through and the voltage over the electrodes. In \emph{galvanostatic} experiments the current is the input and voltage the output, while in \emph{potentiostatic} experiments it is the other way around. In the remainder of this text, EIS experiments satisfying the assumptions of LTI systems will be referred to as `classical EIS'. Estimating transfer functions from input and output data of LTI systems is a thoroughly studied problem in the field of system identification \cite{ljungSystemidentification,pintelon2012system}.

However, for electrochemical systems it is well known that (i) the relation between current and voltage is generally nonlinear, e.g.\ expressed by exponential functions in the Butler-Volmer equation, and (ii) the behaviour of the process may evolve over time. Li-ion batteries, for example, provide such time-variation on different time-scales. On a large time-scale, the impedance of a fresh and an aged cell are different \cite{troltzsch2006characterizing,zhu2018electrochemical}. On a shorter time-scale, the impedance of a fully charged and a fully discharged cell are also different \cite{andre2011characterization}. Accordingly, with classical EIS, one represents electrochemical processes which are by nature nonlinear and nonstationary with a model for linear and stationary systems. While classical EIS may be a non-ideal approximation for such cases, important information about the process is nonetheless revealed by measuring within forced constraints of linearity and stationarity at specific operating points. Linearity is achieved by applying a small excitation amplitude such that the behaviour of the process is linearised in a certain operating region. Stationarity is achieved by measuring at a time, and over a timespan, when the process is and remains in steady-state. However, these experimental conditions are very restrictive. How can we obtain information about the nonlinear behaviour of a system when it is only possible to perform experiments under linear constraints? How can we study a battery \emph{while} it is charging or discharging? How can we study anodising \emph{while} a protective layer is forming?

To relax these limitations, the concept of impedance needs extensions. The required extensions to model systems beyond the linearity and stationarity constraints have been studied in system theory. Nonlinear time-invariant (NLTI)\nomenclature[A]{NLTI}{Nonlinear time-invariant} systems are commonly studied by \emph{Volterra series} \cite{schetzen1980volterra,boyd1985fading}. These are convolution operators that are able to capture dynamical nonlinear behaviour, where the transfer function is extended to so-called generalised transfer functions. Similarly, nonlinear impedances are studied in nonlinear EIS (NLEIS)\nomenclature[A]{NLEIS}{Nonlinear electrochemical impedance spectroscopy} \cite{harting2017nonlinear,fasmin2017nonlinear,murbach2018nonlinear,vidakovic2021nonlinear,zabara2022non,kirk2022nonlinear,schluter2022nonlinear}. Assuming that the nonlinear behaviour of an electrochemical system can be captured by a Volterra series, the behaviour of the process can be split into a linear part and purely nonlinear part \cite{schoukens2005identification}. The model for the linear part is denoted as the best linear approximation (BLA)\nomenclature[A]{BLA}{Best linear approximation} of the system. When the nonlinearities are small enough compared to the linear behaviour, use of the BLA for describing the system is justified.

Nonstationary systems have been studied by extending the transfer function to be a time-varying transfer function to describe a linear time-varying (LTV)\nomenclature[A]{LTV}{Linear time-varying} system \cite{zadeh1950frequency}. The latter is a transfer function depending on both time and frequency, expressing the evolution of the transfer function over time. Similarly, the impedance can be extended to a \emph{time-varying} impedance. As an analogy, the impedance can be considered to be like a photograph where as in the first days of photography---think mid 19th century---the subject should remain stationary during a certain exposure time for accumulating light on a sheet, but the time-varying impedance can be seen as a movie of a subject during an activity.

Over the years, techniques have been developed to \emph{detect} nonstationarities and nonlinearities in measured data, where it was shown that frequency domain identification techniques with multisine excitations are advantageous \cite{van2004electrochemical,van2006electrochemical,van2009advantages,breugelmans2012odd,hallemans2020detection}. Recently, with increasing computation power, different techniques have been developed and refined for unravelling time-varying impedance from data \cite{bond1977line,darowicki2000theoretical,darowicki2003dynamic,sacci2009dynamic,sacci2014dynamic,koster2017dynamic,hallemanstimevarying}. Time-varying impedance data have already successfully been obtained for a wide variety of electrochemical processes, including organic coatings \cite{wouters2021monitoring}, electrorefining \cite{Collet2021Thiourea}, hydrogen evolution reactions \cite{koster2019extracting}, nickel hexacyanoferrate thin films \cite{erinmwingbovo2019dynamic}, electrochemical double layer capacitors \cite{pianta2022evaluation}, charging/discharging Li-ion batteries \cite{huang2014exploring,huang2015dynamic,zappen2018application,zhu2022operando,hallemans2022operando}, and Li-plating in batteries \cite{brown2021detecting,koseoglou2021lithium,katzer2021analysis}.

Commonly, drift signals may appear (for instance the slow voltage increase when charging a battery) during \emph{in operando} electrochemical measurements. These drift signals prohibit measuring impedance data at low frequencies. A method for removing drift signals has been developed \cite{hallemans2022TrendRemoval}, and successfully applied to electrorefining \cite{collet2022time}, anodising \cite{havigh2022operando}, and Li-ion batteries \cite{zhu2022operando,hallemans2022operando}, such that the time variations of the low-frequency impedance can also be estimated.

In this review article, we detail the required mathematical concepts associated with impedance and their extensions beyond linearity and stationarity. Key concepts are supported with illustrations obtained from simulations and real-life measurements. Experiments are performed on a pristine commercially available Samsung 48X Li-ion battery. This is a $4.8\,\mathrm{Ah}$ $21$ $700$ cylindrical cell format with cathodes based on lithiated metal oxide (Co, Ni, Al) and anodes based on graphite and blended Si. The impedance is measured at different temperatures, while resting and while charging. We opt for this case-study on Li-ion batteries since EIS is becoming a popular tool for characterising batteries, diagnosing state-of-health, and developing smart charging protocols. These compelling applications are also discussed as a motivation to perform EIS beyond linearity and stationarity. Moreover, the measurements are obtained using a commercial potentiostat (Gamry Interface 5000E), showcasing the practical accessibility of the discussed modelling techniques. 

This article is structured as follows. First, we give a motivational example on how impedance data beyond linearity and stationarity is promising for battery aging diagnostics and smart charging protocols (Section~\ref{Section:whyEIS}). Next, we define what we mean with a model for the electrochemical system (Section~\ref{Section:modellingElectrochemicalSystems}). Then, we revisit classical EIS (Section~\ref{Section:classicalEIS}), with emphasis on the limiting constraints of linearity and stationarity. The choice between single-sine and multisine excitations is discussed in depth. Then, we formally introduce nonlinear and nonstationary models for electrochemical measurements through, respectively, Volterra series and time-varying impedances (Section~\ref{Section:NLandNSModels}). The Volterra series is linked to NLEIS and the BLA. Next, we detail the experimental procedure in measuring current and voltage time-series for proper impedance measurements (Section~\ref{Section:DataCollection}). The estimation of classical impedance data in the frequency domain from periodic and random excitations is discussed in Section~\ref{Section:classicalImpedanceEstimationFD}. Then, we detail how the linearity and stationarity constraints can be assessed, and nonlinearities and nonstationarities are detected by observing the current and voltage spectra under odd random phase multisine excitations (Section~\ref{Section:DetectionNLNS}). Obtaining nonlinear impedance data and the BLA is discussed in Section~\ref{Section:NonlinearModelEstimation}. The extraction of time-varying impedance data from the collected current and voltage data is studied through different relevant methods in Section~\ref{Section:timeVaryingImpedanceEstimation}. This has been studied in Szekeres et al.\ \cite{szekeres2021methods} also, however, here a deeper mathematical foundation is given. In Section~\ref{Section:caseStudyLiIon}, the performed illustrative experiments on Li-ion batteries are discussed. Finally, conclusions are drawn and an outlook is given in Section~\ref{Section:conclusions}.

\section{Overview of applications of EIS for batteries}\label{Section:whyEIS}
Before we get into the details of how impedance works, let us first motivate the topic and reflect on why impedance is useful for solving some of electrochemistry's crucial research problems, and more importantly, why measuring impedance beyond linearity and stationarity is promising. We do this for the compelling case of Li-ion batteries. Here, some relevant research problems include state-of-health (SOH)\nomenclature[A]{SOH}{State-of-health} prognostics \cite{zhang2011review,waag2014critical,berecibar2016critical,birkl2017degradation,xiong2018towards,li2019data,pastor2019critical,severson2019data,o2022lithium,attia2022knees} and smart charging \cite{tomaszewska2019lithium,attia2020closed,couto2021faster,wang2022fast}. For SOH prognostics, EIS is a powerful non-invasive tool \cite{hu2023application}. It has been shown that classical impedance data, mapped onto equivalent circuit model (ECM)\nomenclature[A]{ECM}{Equivalent circuit model} parameters, contains important information about degradation mechanisms \cite{pastor2017comparison,pastor2019critical,jiang2022comparative}. Moreover, classical impedance data is an informative input to machine learning algorithms to predict the remaining-useful-life (RUL)\nomenclature[A]{RUL}{Remaining-useful-life} of batteries \cite{zhang2020identifying,gasper2022predicting}. Jones et al.\ \cite[Table 1]{jones2022impedance} demonstrate that the EIS data state representation performs better than other state representations for SOH forecasting. Bizeray et al.\ \cite{bizeray2018identifiability} also show that physical parameters can be identified from classical impedance data, allowing simulation of the battery using physics-based models such as the single-particle model (SPM)\nomenclature[A]{SPM}{Single-particle model}. This is an non-invasive way of parametrising cells, being more practical than tearing down cells \cite{chen2020development}.

Impedance data beyond linearity and stationarity has the potential to improve battery SOH diagnostics since it contains additional information compared to classical impedance data. Leveraging nonlinear and time-varying impedance data has already been carried out for detecting Li-plating \cite{harting2018identification,katzer2021analysis} (an important degradation mechanism in Li-ion batteries \cite{birkl2017degradation,edge2021lithium}). Moreover, Kirk et al.\ \cite{kirk2022nonlinear} demonstrate that nonlinear impedance data contributes to the identifiability of physical SPM parameters.

Impedance data beyond linearity and stationarity also has the potential to improve smart charging protocols. Katzer et al.\ \cite{katzer2023adaptive} propose an adaptive fast charging protocol relying on impedance-based detection of Li-plating \cite{katzer2021analysis}. In Zhu et al.\ \cite{zhu2022operando}, we track the charge transfer resistance while charging, which can be obtained from time-varying impedance data, and propose to adapt the charge profile based on this time-varying charge transfer resistance.

A critical reader may argue that EIS experiments are expensive, and cannot easily be implemented in battery management systems (BMS)\nomenclature[A]{BMS}{Battery management system}. However, different solutions have been developed to implement low-cost measurement apparatus \cite{howey2013online,olarte2021battery,carbone2022low}. Here multisine excitations and frequency domain model estimation are promising tools.

\section{Modelling electrochemical systems}\label{Section:modellingElectrochemicalSystems}
Electrochemical processes are often studied by modelling the relation between the current $i(t)$\nomenclature[F]{$i(t)$}{Current signal} through, and the voltage $v(t)$\nomenclature[F]{$v(t)$}{Voltage signal} over the electrodes, where $t$\nomenclature[V]{$t$}{Time} denotes continuous time. However, this relation may also be dependent on external parameters $p(t)$\nomenclature[F]{$p(t)$}{External parameters} such as the ambient temperature, the rotation rate of the electrodes, the pressure, the concentration distribution on the electrodes, etc. 

Two types of experiments are common to measure the relation between current and voltage. In \emph{galvanostatic} experiments, a current is applied and the voltage is measured. In a general setting, the impedance is modelled as an operator $\mathcal{G}\{\cdot\}$ acting on the current and external parameters,
\begin{align}
v(t)=\mathcal{G}\{i(t),p(t)\}.
\label{eq:operator}
\end{align}
In \emph{potentiostatic} experiments, the measurements are performed the other way around, that is, through an operator $\mathcal{P}$, $i(t)=\mathcal{P}\{v(t),p(t)\}$. Here, the notations for galvanostatic experiments are chosen, that is, the current $i(t)$ is the excitation and the voltage $v(t)$ is the response (appropriate for low impedance devices such as batteries).

As mentioned in the introduction, the operators above are dynamical, that is, operators with memory, or also called \emph{convolution operators}, not only acting on the present time, but also making use of past information. 

In what follows, it is detailed how the operator $\mathcal{G}$ can be modelled by an impedance. We first consider the system under the constraints of linearity and stationarity (Section~\ref{Section:classicalEIS}), and proceed by introducing models beyond these hard constraints (Section~\ref{Section:NLandNSModels}).

\section{Classical EIS revisited}\label{Section:classicalEIS}
\begin{figure}
    \centering
    \includegraphics[width=0.4\textwidth]{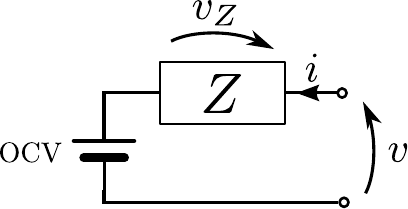}
    \caption{Equivalent electrical schematic of an electrochemical system under LTI constraints.}
    \label{fig:LTImodelSchematic}
\end{figure}
\subsection{The constraints of classical EIS}\label{Section:constraintsClassicalEIS}
In classical EIS experiments, the external parameters $p(t)$ are assumed constant during the experiment and the generic model \eqref{eq:operator} is simplified to
\begin{align}
v(t)=\mathrm{OCV}+\underbrace{Z\{i(t)\}}_{v_Z(t)},
\end{align}
where OCV\nomenclature[A]{OCV}{Open circuit voltage} is the open circuit voltage, assumed constant, $Z$ is the classical impedance operator and $v_Z(t)$\nomenclature[F]{$v_Z(t)$}{Voltage over the impedance} is the voltage over the impedance. An equivalent electrical circuit representing this model is shown in Fig.~\ref{fig:LTImodelSchematic}. It is assumed that operator $Z$ satisfies the constraints of LTI systems, that is, linearity and stationarity, and also causality. 
\paragraph{Linearity} The operator $Z$ is a linear operator, satisfying additivity and homogeneity, respectively,
\begin{subequations}
\begin{align}
& Z\{i_1(t)+i_2(t)\}=Z\{i_1(t)\}+Z\{i_2(t)\}\label{eq:additivity} \\
& Z\{\alpha i(t)\}=\alpha Z\{i(t)\} \quad \forall \alpha \in \mathbb{R}.
\end{align}
\label{eq:linearity}
\end{subequations}
As such, when the current $i$ doubles, the voltage $v_Z$ over the impedance also doubles.
\paragraph{Stationarity (time-invariance)} A stationary system is a system whose behaviour does not change when shifted in time. Accordingly, the operator $Z$ is independent of the time at which the excitation is applied: 
\begin{align}
Z\{i(t-\tau)\}= v_Z(t-\tau) \quad \forall \tau\in \mathbb{R}.
\label{eq:stationarity}
\end{align}
\paragraph{Causality} The response of the system is totally determined by the excitation. As a consequence, the response to an excitation cannot precede the excitation.

For \emph{potentiostatic} experiments under LTI constraints, the excitation-response relation yields,
\begin{align}
i(t)=Y\{\underbrace{v(t)-\mathrm{OCV}}_{v_Z(t)}\},
\end{align}
where $Y$ is called the admittance operator, satisfying the same conditions as the impedance operator $Z$. Notations in this paper can, hence, be converted into potentiostatic experiments by swapping $i$ and $v_Z$, and replacing the impedance $Z$ by admittance $Y$.

\begin{figure}
    \centering
    \includegraphics[width=0.5\textwidth]{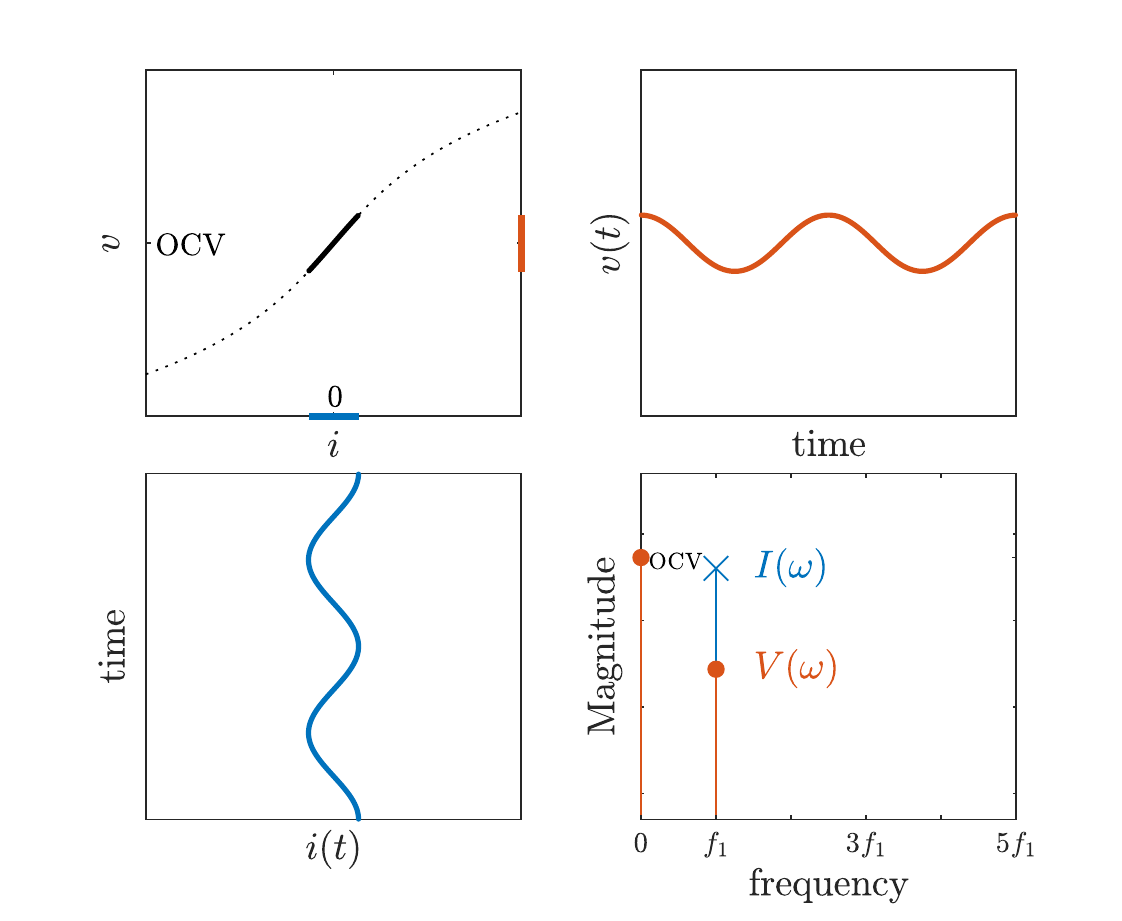}
    \caption{Illustration of the linearisation of the Butler-Volmer equation \eqref{eq:vZasFunctionOfiBV} for a small amplitude sinusoidal excitation.}
    \label{fig:nonlinearity1}
\end{figure}

When charge transfer is the rate-determining step, the static relation between current and voltage of electrochemical reactions is described by the Butler-Volmer equation \cite{dickinson2020butler},
\begin{align}
i=j_0 S\left(\exp\left(\frac{\alpha_a n F}{R\mathrm{T}}v_\mathrm{Z}\right)-\exp\left(-\frac{\alpha_c n F}{R\mathrm{T}}v_\mathrm{Z}\right)\right),
\label{eqn:BV}
\end{align}
with $j_0$ the exchange current density, $S$ the surface area of the electrode, $\mathrm{T}$ the absolute temperature in Kelvin, $n$ the number of electrons, $F$ the Faraday constant, $R$ the universal gas constant, $\alpha_a$ the anodic charge transfer coefficient, and $\alpha_c$ the cathodic charge transfer coefficient. When assuming $\alpha_a=\alpha_c=0.5$, this equation can be rewritten to obtain the overpotential as a function of the current,
\begin{align}
    v_\mathrm{Z}=\frac{2R\mathrm{T}}{nF}\sinh^{-1}\left(\frac{i}{2j_0S}\right).
    \label{eq:vZasFunctionOfiBV}
\end{align}
The top left plot of Fig.~\ref{fig:nonlinearity1} shows this Butler-Volmer relation between current and voltage as a dotted line. This relation is obviously not linear. However, the linearity constraint can approximately be satisfied by choosing the magnitude of the excitation signal in a specific range. This can be seen by expanding \eqref{eq:vZasFunctionOfiBV} as a Taylor series around $i=0$,
\begin{align}
    v_\mathrm{Z}= \underbrace{\frac{\partial v_\mathrm{Z}}{\partial i}\bigg\rvert_{i=0}i}_\text{linear term}+\frac{\partial^2 v_\mathrm{Z}}{\partial i^2}\bigg\rvert_{i=0}i^2+\frac{\partial^3 v_\mathrm{Z}}{\partial i^3}\bigg\rvert_{i=0}i^3+\hdots
    \label{eq:TaylorSeriesBV}
\end{align}
When the current $i$ is small enough, the linear term will dominate the higher order terms and linearity can be assumed. A rule of thumb for ensuring linearity is that the voltage deviation should not be larger than \SI{15}{mV} \cite{macdonald2018impedance}. An illustration of the linearisation of the Butler-Volmer equation is also shown in Fig.~\ref{fig:nonlinearity1}. A small amplitude sinusoidal current excitation centered around zero (blue) is applied to the Butler-Volmer equation (dotted black line), and the response is the voltage centered around the value OCV (red). For an excitation with small amplitude, the Butler-Volmer equation is quasi-linear in the excited range (black line).

The stationarity constraint, on the other hand, is satisfied by driving the electrochemical system in steady-state, and applying a zero-mean current excitation to remain in steady-state. For a battery, for instance, applying a current excitation with a positive mean value would charge the battery and cause nonstationary behaviour. It is also necessary that the external parameters $p(t)$ remain constant during the experiment. A significant change in ambient temperature, for instance, might jeopardise the stationarity constraint.

The relation between the excitation and response of LTI systems is well documented in system theory  \cite{kailath1980linear}. The response $v_\mathrm{Z}(t)$ of an LTI system is commonly modelled by the \emph{convolution} of the impulse response function $z(t)$ with the excitation $i(t)$, 
\begin{align}
    v_Z(t)=\int_{-\infty}^\infty z(\tau)i(t-\tau)\mathrm{d}\tau.
    \label{eq:vconvTimeDomain}
\end{align}
The impulse response function\nomenclature[F]{$z(t)$}{Impulse response function} is the response to a Dirac pulse. Note that \eqref{eq:vconvTimeDomain} satisfies \eqref{eq:linearity} and \eqref{eq:stationarity}. Often, the fact that a convolution in the time domain becomes a product in the frequency domain is exploited to rewrite \eqref{eq:vconvTimeDomain} in a way where the frequency dependent impedance $Z(\omega)$\nomenclature[F]{$Z(\omega)$}{Impedance}\nomenclature[F]{$Z(\omega,t)$}{Time-varying impedance} appears,
\begin{align}
v_Z(t)=\mathcal{F}^{-1}\{Z(\omega)I(\omega)\}.
\label{eq:vtZwIW_LTI}
\end{align}
Here, the impedance $Z(\omega)$ is defined as the Fourier transform of the impulse response function $z(t)$, $\mathcal{F}^{-1}\{\cdot\}$\nomenclature[O]{$\mathcal{F}\{\cdot\}$}{Fourier transform}\nomenclature[O]{$\mathcal{F}^{-1}\{\cdot\}$}{Inverse Fourier transform} is the inverse Fourier transform operator, and $I(\omega)$ the Fourier transform of the current. The Fourier and inverse Fourier transforms are defined in \ref{app:FT}. Recall that the angular frequency $\omega$ is related to the frequency $f$ as $\omega=2\pi f$\nomenclature[V]{$\omega$}{Angular frequency}\nomenclature[V]{$f$}{Frequency}.

The voltage over the electrochemical system is then
\begin{align}
v(t)=\mathrm{OCV}+\mathcal{F}^{-1}\{Z(\omega)I(\omega)\}.
\end{align}
Accordingly, the impedance is given by the \emph{ratio of the Fourier transforms of voltage and current},
\begin{align}
Z(\omega)=\frac{V(\omega)}{I(\omega)} \qquad \omega\neq0.
\label{eq:ZomegaSimpleCont}
\end{align}
Note that the impedance is not expressed at DC. The impedance at DC becomes infinite in magnitude and with a purely imaginary phase because the linearised OCV behaves like a capacitor.

As for any frequency response function, the impedance is a complex valued function, often denoted as
\begin{align}
Z(\omega)=Z_\mathrm{r}(\omega)+jZ_\mathrm{j}(\omega),
\end{align}
\nomenclature[F]{$Z_\mathrm{r}$}{Real part of the impedance}
\nomenclature[F]{$Z_\mathrm{j}$}{Imaginary part of the impedance}
\nomenclature[C]{$j$}{Imaginary unit}
with $j$ the imaginary unit ($j^2=-1$), and $Z_\mathrm{r}(\omega)$ and $Z_\mathrm{j}(\omega)$ the real and imaginary parts of the impedance, respectively. The complex-valued impedance is also defined by its \emph{magnitude} and \emph{phase}, respectively,
\begin{subequations}
\begin{align}
\vert Z(\omega)\vert &=\sqrt{Z_\mathrm{r}^2(\omega)+Z_\mathrm{j}^2(\omega)}\\
\angle Z(\omega)&=\left\{
    \begin{array}{ll}
         \arctan \frac{Z_\mathrm{j}(\omega)}{Z_\mathrm{r}(\omega)} &  \text{for } Z_\mathrm{r}(\omega)\geq 0\\
        \pi+\arctan \frac{Z_\mathrm{j}(\omega)}{Z_\mathrm{r}(\omega)} & \text{for } Z_\mathrm{r}(\omega)<0.
    \end{array}
\right.
\end{align}
\end{subequations}
\nomenclature[O]{$\vert \cdot \vert$}{Magnitude}
\nomenclature[O]{$\angle \cdot$}{Phase ($\varphi$)}
Note that in electrochemistry the phase of the impedance is often denoted as $\varphi(\omega)$.

\begin{figure}
    \centering
    \includegraphics[width=0.5\textwidth]{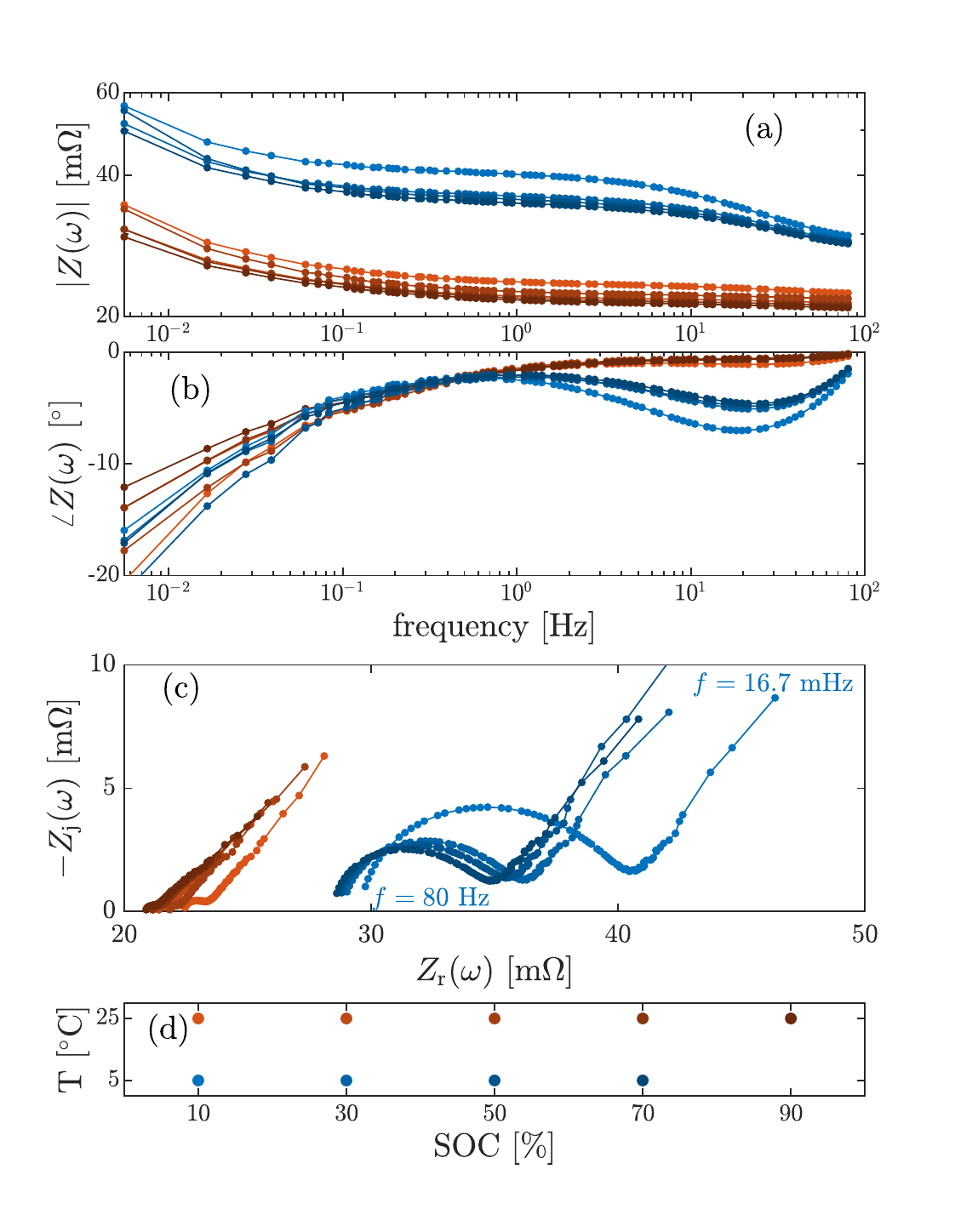}
    \caption{Illustration of classical impedance data at different operating points for a Samsung 48X Li-ion battery (see Section~\ref{Section:caseStudyLiIon}). The operating points in the temperature-SOC plane are shown in (d), while the corresponding impedances are shown as Bode plot in (a-b) and as Nyquist chart in (c). The impedance data at the selected frequencies is indicated by dots, which are connected with straight lines.}
    \label{fig:cEISOCV}
\end{figure}

The impedance is usually visualised on a Bode plot (Fig.~\ref{fig:cEISOCV} (a-b)) as magnitude and phase in function of frequency, or on a Nyquist chart (Fig.~\ref{fig:cEISOCV} (c)) as real versus \emph{negative} imaginary part, since electrochemical systems are often capacitative in their electrical behaviour.

In the \emph{potentiostatic} case, the impedance can still be computed by \eqref{eq:ZomegaSimpleCont}, since the admittance $Y(\omega)$ is defined as follows,
\begin{align}
Y(\omega)=\frac{I(\omega)}{V(\omega)}=\frac{1}{Z(\omega)}.
\end{align}
\nomenclature[F]{$Y$}{Admittance}
\subsubsection*{Kramers-Kronig} The conformity of the required constraints of linearity \eqref{eq:linearity} and stationarity \eqref{eq:stationarity} can be validated through the Kramers-Kronig transformation \cite{wang2021electrochemical,urquidi1986application,urquidi1990applications,boukamp1995linear}. This transformation states that there is an analytical relation between the real and imaginary parts of the impedance,
\begin{subequations}
\begin{align}
Z_\mathrm{r}(\omega)&=Z_\mathrm{r}(\infty)+\frac{2}{\pi} \int_0^\infty \frac{xZ_\mathrm{j}(x)-\omega Z_\mathrm{j}(\omega)}{x^2-\omega^2}\mathrm{d}x\\
Z_\mathrm{j}(\omega)&=-\frac{2\omega}{\pi} \int_0^\infty \frac{Z_\mathrm{r}(x)-Z_\mathrm{r}(\omega)}{x^2-\omega^2}\mathrm{d}x.
\end{align}
\label{eq:Kramers-Kronig}
\end{subequations}
In theory, when the imaginary impedance computed from the measured real part coincides well with the measured imaginary part, or vice versa, the required constraints are assumed to be satisfied. However, in practice, \eqref{eq:Kramers-Kronig} is difficult to implement since the integral needs continuous impedance data (while only discrete data is available) and goes from DC to infinity (while data is only available in a certain frequency band). Moreover, measurement noise is not accounted for. Hence, an alternative approach is to fit an equivalent circuit model based on solely the real, or imaginary, part of the impedance data. When the model coincides well with the measured complex impedance data, the Kramer-Kronig relation is assumed to be satisfied \cite{agarwal1995application,roy2007error}.

\subsubsection*{Models at operating points}
It is very important to stress that the impedance $Z(\omega)$ measured with classical EIS is dependent on the local operating point (in the sense of a Taylor expansion) at which the experiments have been performed. The operating point is defined by the OCV value and the constant values of the external parameters $p(t)$. As an example, Fig.~\ref{fig:cEISOCV} (d) shows operating points of a Li-ion battery depending on the state-of-charge (SOC)\nomenclature[A]{SOC}{State-of-charge} and temperature. The measured classical impedances $Z(\omega)$ of a Samsung 48X Li-ion battery (see Section~\ref{Section:caseStudyLiIon}) at these operating points are shown as Bode (a-b) and Nyquist (c) plots. We observe that the impedance depends on the SOC and temperature, and that low SOC and low temperature exhibit higher impedance values. Furthermore, for batteries, a difference in impedance would also be visible for experiments at the same SOC and temperature, but at different SOH \cite{troltzsch2006characterizing,zhu2018electrochemical}.

\subsection{Excitation}\label{Section:ExcitationSignalsClassicalEIS}
The excitation signal $i(t)$ must be `rich' enough such that the response $v(t)$ contains the information needed for extracting impedance data. Since impedance data should be measured at a set of frequencies, it is natural to use sinusoidal functions as excitations. Historically, EIS was performed with single-sine excitations. Later, multisine EIS was developed \cite{van2009advantages}. Both of them have pros and cons, which are discussed now.
\begin{figure}
    \centering
    \includegraphics[width=0.5\textwidth]{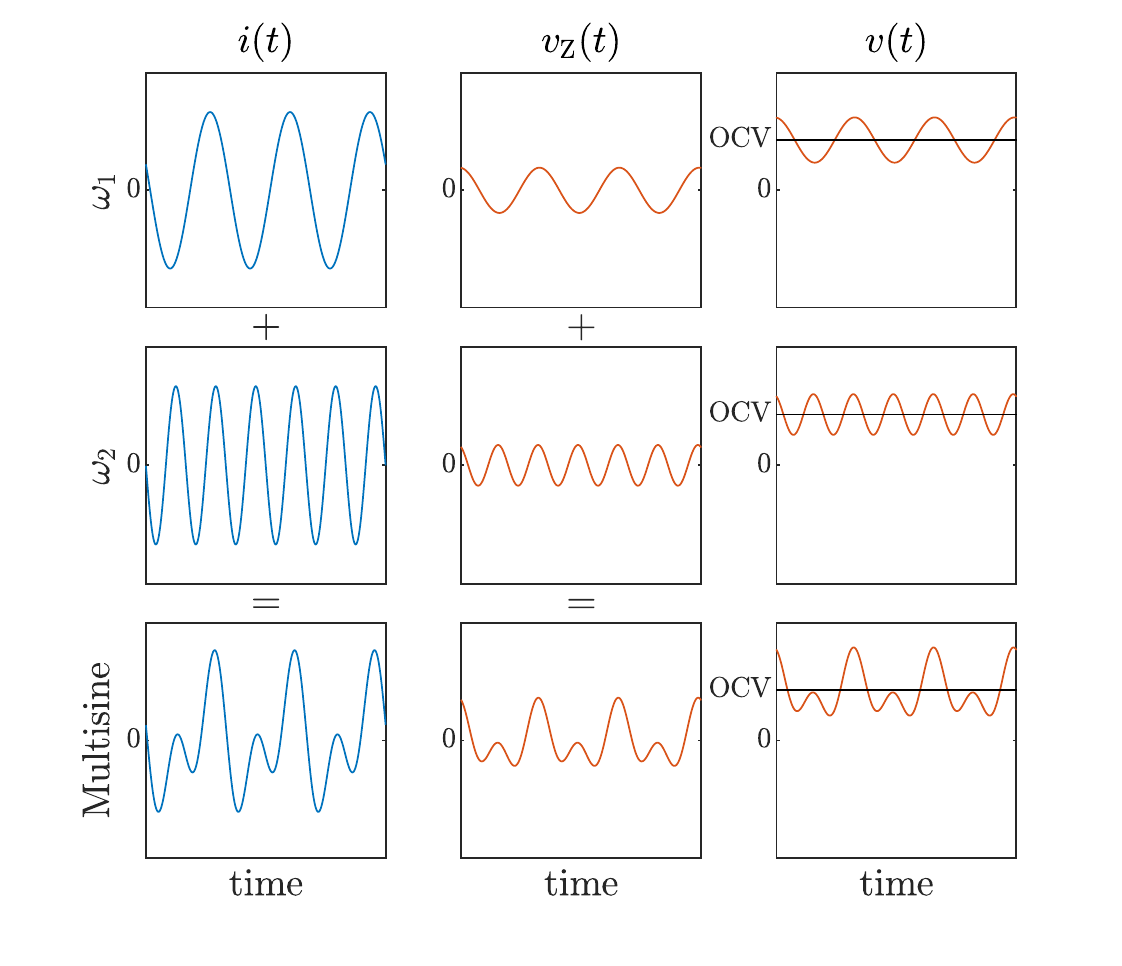}
    \caption{Illustration of the response of an LTI system excited by single-sine and multisine excitations. The first row has a single-sine excitation at angular frequency $\omega_1$, the second row at $\omega_2=2\omega_1$, and the third row has a multisine excitation, which is the sum of the two single-sines. The OCV is plotted in black.}
    \label{fig:ivSingleSineMultisine}
\end{figure}
\subsubsection{Single-sine excitation}
Commonly, single-sine excitations are used for EIS \cite{BiologicEIS,GamryEIS}. That is, a sinusoidal zero-mean current signal with small amplitude $I_m$ at a selected angular frequency $\omega_m$ is applied,
\begin{align}
i(t)= I_m\cos(\omega_m t),
\label{eq:singleSine}
\end{align}
and the voltage response \eqref{eq:vtZwIW_LTI} is measured,
\begin{align}
v(t)=\mathrm{OCV}+\underbrace{\vert Z(\omega_m)\vert I_m}_{V_m}\cos\big(\omega_m t+\underbrace{\angle Z(\omega_m)}_{\psi_m}\big).
\label{eq:responseSingleSine}
\end{align}
The voltage response is a sinusoidal signal (assuming linearity due to the small current amplitude) at the same frequency, however, with a different amplitude $V_m$ and phase $\psi_m$, superimposed on the OCV. This is illustrated in Fig.~\ref{fig:nonlinearity1} and the two top rows of Fig.~\ref{fig:ivSingleSineMultisine}. The complex-valued impedance at angular frequency $\omega_m$ is computed from the amplitude scaling and phase shift between current and voltage,
\begin{align}
Z(\omega_m)=\frac{V_m}{I_m}e^{j\psi_m}.
\label{eq:impedanceLTI1}
\end{align}
The selected frequencies $\omega_m$, $m=1,2,...,M$, are applied \emph{sequentially} (i.e.\ one after the other), usually starting from the highest frequency and ending at the lowest one. Often the selected frequencies are logarithmically spaced over multiple decades such as to excite processes happening at different time-scales. The impedance at each of these frequencies is computed. Note that it is only possible to apply the sinusoids sequentially because stationarity \eqref{eq:stationarity} is assumed, and, hence, the response is independent of the time of excitation. 
\subsubsection{Multisine excitation}
Instead of applying sinusoidal signals \emph{sequentially}, it is also possible to apply the different frequencies \emph{simultaneously}. This is the purpose of a multisine, where the \emph{sum} of sinusoidal signals at different frequencies is applied,
\begin{align}
i(t)= \sum_{m=1}^M I_m\cos(\omega_m t+\phi_m).
\label{eq:multisine1}
\end{align}
Here, each of the sinusoidal components is given a different phase $\phi_m$ such as not to introduce constructive interference (see Section~\ref{Section:DataCollection}). Due to the linearity constraint \eqref{eq:linearity}, the total response yields the sum of each of the individual responses,
\begin{align}
v(t)=\mathrm{OCV}+\sum_{m=1}^M \underbrace{\vert Z(\omega_m)\vert I_m}_{V_m}\cos\big(\omega_m t+\underbrace{\phi_m+\angle Z(\omega_m)}_{\psi_m}\big).
\label{eq:responseMultisine}
\end{align}
The impedance values at the selected angular frequencies $\omega_m$ can still be computed based on the amplitude scaling and the phase shifts of the sinusoidal components,
\begin{align}
Z(\omega_m)=\frac{V_m}{I_m}e^{j(\psi_m-\phi_m)}.
\label{eq:impedanceLTI}
\end{align}
An illustration of current and voltage signals for a multisine excitation under the assumptions of linearity and stationarity is shown in the bottom row of Fig.~\ref{fig:ivSingleSineMultisine}.

\subsubsection{The choice of excitation signal}\label{Section:theChoiceOfExcitationSignal}
For classical EIS experiments, it is common to use single-sine experiments since they are known to electrochemists and easy to use with commercially available potentiostats. However, for broadband
experiments, that is, experiments over a large frequency band, we recommend the use of a multisine. 

First of all, since for multisine experiments all frequencies are applied \emph{simultaneously}, as opposed to \emph{sequentially} in single-sine experiments, the experiment time is shorter \cite{van2009advantages}. Moreover, in single-sine experiments we should wait for transients to fade out at each individual frequency, while for multisine experiments this should only be done once, which also decreases the experiment time. 

The constraints of linearity and stationarity are easily checked for multisine experiments by looking at the measured current and voltage data in the frequency domain (see Section~\ref{Section:DetectionNLNS}), while for single-sine experiments one needs to check the Kramers-Kronig relations. It is noteworthy that for multisine experiments one must use the frequency domain to detect nonlinearity and nonstationarity, since it has been empirically shown that multisine experiments always satisfy the Kramers-Kronig relations \cite{you2020application}. This is studied in more detail in Section~\ref{Section:DetectionNLNS}. Next, we demonstrate that the most commonly stated arguments in favor of single-sine experiments over multisine ones can be debunked. 
\begin{enumerate}
\item The linearity constraint is believed to be easier to impose in the single-sine setting. The amplitudes such that the response stays linear can be determined for each of the sines separately. In the multisine case, since a high number of frequencies are added, the amplitude of each frequency separately should remain small such that the total multisine signal does not become too high in magnitude (in a root-mean-square sense) and introduce nonlinearities. However, finding the optimal excitation amplitudes for multisine experiments can also be done by detecting nonlinear behaviour. Moreover, the amplitudes can be dependent on the frequency too.
\item The signal-to-noise ratio (SNR) is believed to be better in the single-sine case. This is since all the power is injected at one frequency, while, in the multisine case, there is a trade-off between the number of selected frequencies and the SNR. The more frequencies are excited, the smaller their amplitudes must be for the response to remain linear, and hence, the smaller the SNR. However, it is important to take the measurement time into account, and compare the SNR for single-sine and multisine experiments of the same duration \cite[Section 5.2.2 p. 154]{pintelon2012system}. Since single-sine experiments take a longer time, we can measure more periods of the multisine during the same measurement time, resulting in a better SNR.
\item For a multisine excitation, the selected frequencies should all be integer multiples of a fundamental frequency such that the period of the multisine equals the period of the fundamental sine. These integer multiples are called \emph{harmonics}. Single-sine experiments do not have this limitation. Accordingly, multisine experiments have less flexibility in the choice of frequencies at the low frequency bands. However, this is not a significant issue; one could perform multiple multisine experiments with slightly different fundamental frequencies if required.
\item The extraction of the impedance in the single-sine case is very intuitive, and can be handled in the time domain. For multisine excitations, the Fourier transform is usually used for separating the frequency components, and the impedance is computed by a ratio of Fourier spectra (see Section~\ref{Section:classicalImpedanceEstimationFD}). Working in the time domain only, however, is limited because one might not see leakage, nonlinear distortions and nonstationarity.
\end{enumerate}
Another strong argument in favour of multisine experiments is that if the system behaves in a nonstationary way, one \emph{should} use a multisine excitation, as pointed out in Section~\ref{Section:Nonstationarity}.

In Section~\ref{Section:classicalImpedanceEstimationFD}, we study how classical impedance data can be estimated from measured current and voltage data.

\section{Nonlinear and nonstationary impedance models}\label{Section:NLandNSModels}
In this section, impedance models are introduced beyond the very restrictive constraints of linearity and stationarity. Eliminating the linearity constraint leads to the Volterra series model, from which the concepts of nonlinear EIS (NLEIS) and the best linear approximation (BLA) are derived (Section~\ref{section:NonlinearModels41}). Getting rid of the stationarity constraint leads to the time-varying impedance model (Section~\ref{Section:Nonstationarity}). Eliminating both the constraints of linearity and stationarity, we will study the best linear time-varying approximation (BLTVA)\nomenclature[A]{BLTVA}{Best linear time-varying approximation} (Section~\ref{section:nonlinearAndNonstationaryModels}). 

\subsection{Nonlinear models}\label{section:NonlinearModels41}
\begin{figure}
    \centering
    \includegraphics[width=0.5\textwidth]{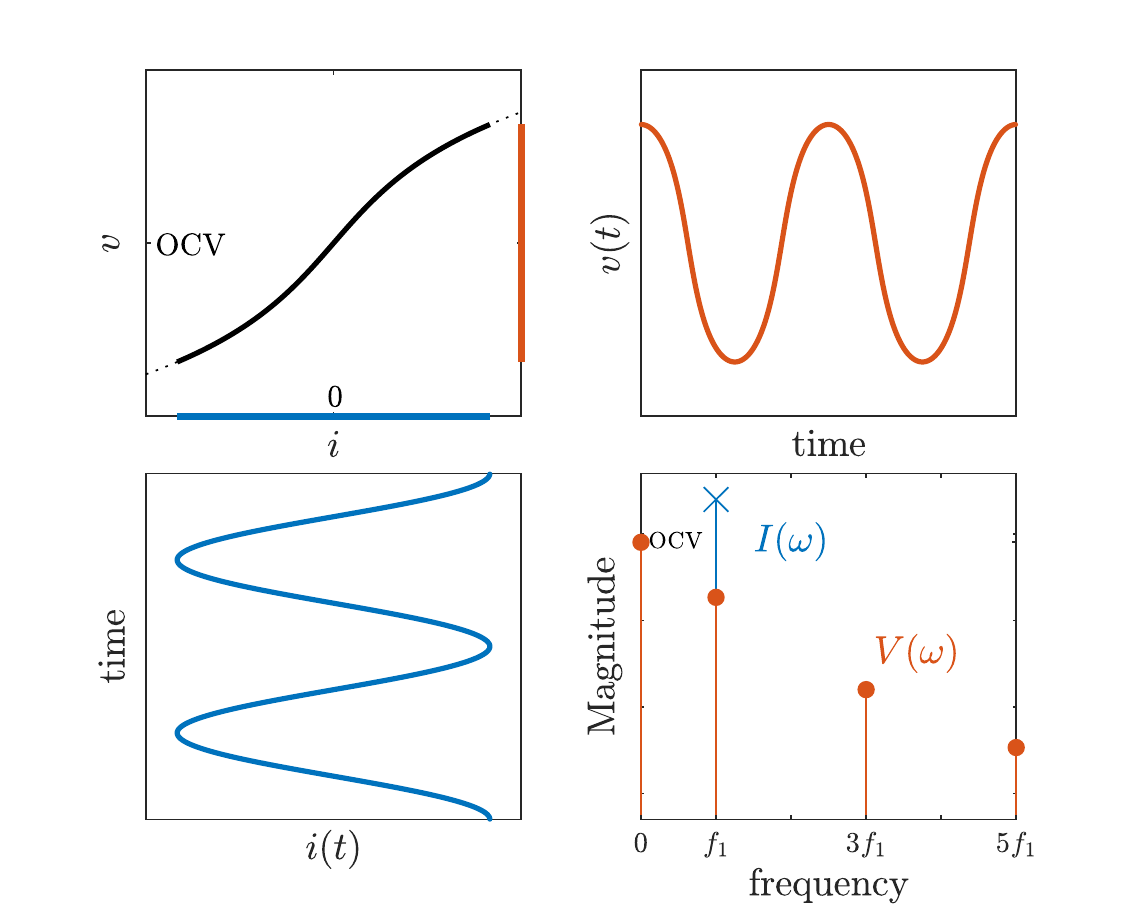}
    \caption{Illustration of nonlinear time-invariant behaviour in EIS measurements under a large excitation amplitude. The simulated nonlinearity comes from the Butler-Volmer equation \eqref{eq:vZasFunctionOfiBV}.}
    \label{fig:nonlinearity2}
\end{figure}

The choice of the excitation magnitudes $I_m$ in classical EIS measurements is a compromise between the need to achieve linearity and the need for a sufficient signal-to-noise ratio \cite{wang2021electrochemical}. EIS measurements, just as any other measurements, contain noise. This noise is caused by external disturbances and the electronics of the measurement device. To improve the quality of impedance data, one can increase the magnitudes $I_m$ in the excitation such that the voltage response becomes more dominant over the noise level. However, this may jeopardise the assumption of linearity since the amplitude span over which the current or voltage is perturbed increases (see Fig.~\ref{fig:nonlinearity2}). In the case where weak nonlinear distortions are present in measurements, the estimation of the best linear approximation (BLA) \cite{schoukens2005identification} is a convenient tool. On the other hand, we could also intentionally introduce nonlinear behaviour in experiments to investigate additional properties of the electrochemical system, as studied in NLEIS \cite{fasmin2017nonlinear,murbach2018nonlinear,zabara2022non,kirk2022nonlinear} or intermodulated differential immittance\footnote{Impedance and admittance as a combined concept.} spectroscopy \cite{battistel2013nonlinear,battistel2015intermodulated1,battistel2015intermodulated2}. In this section, we introduce a nonlinear model for EIS measurements through the Volterra series, derive NLEIS from it, and study the concept of BLA. It is noteworthy that these are still models at fixed local operating points, though, but valid over a larger excitation amplitude than classical EIS measurements.
\subsubsection{The Volterra series}
Nonlinear time-invariant (NLTI) systems can often be modelled by Volterra series \cite{schetzen1980volterra,boyd1985fading,lang1996output}. Note that this is not the case for all NLTI systems. Systems with subharmonics or chaotic behaviour, for instance, cannot be captured by Volterra series. Fortunately, most electrochemical systems can. Since stationarity is still assumed and we are looking at nonlinear behaviour at specific local operating points, the current excitation signal should be zero-mean and the external parameters constant. Mathematically, the general relation between output voltage and input current \eqref{eq:operator} may be written as
\begin{subequations}
\begin{align}
v(t)=\mathrm{OCV}+\sum_{n=1}^{n_\mathrm{max}} v_n(t),
\end{align}
with $n_\mathrm{max}$ the order of the nonlinearity, with
\begin{align}
v_n(t)=\int_{-\infty}^\infty\cdots \int_{-\infty}^\infty z_n(\tau_1,...,\tau_n)\prod_{l=1}^ni(t-\tau_l)\mathrm{d}\tau_l
\end{align}
\label{eq:VolterraResponsen}
\end{subequations}

\noindent being the contribution of the $n$-th order nonlinearity to the voltage output signal, and $z_n(\tau_1,\hdots,\tau_n)$ the generalised impulse response of the $n$-th order nonlinearity. 

The Volterra series can be understood as the extension of a Taylor expansion \eqref{eq:TaylorSeriesBV} at a local operating point for dynamical systems. The nonlinear behaviour is written as a polynomial (instead of linear) operator acting on the excitation current.

The generalised impedances $Z_n$ are defined as the $n$-dimensional Fourier transform of the generalised impulse responses $z_n$:
\begin{align}
&Z_n(\omega_1,...,\omega_n)=\int_{-\infty}^\infty \cdots \int_{-\infty}^\infty z_n(\tau_1,\hdots,\tau_n)\prod_{l=1}^ne^{-j\omega_l\tau_l}\mathrm{d}\tau_l.
\end{align}
For nonlinearity order $n_\mathrm{max}=1$, the current-voltage relations of linear systems \eqref{eq:vconvTimeDomain} and \eqref{eq:vtZwIW_LTI} are retrieved from the Volterra series,
\begin{align}
v_1(t)&=\int_{-\infty}^\infty z_1(\tau)i(t-\tau)\mathrm{d}\tau=\mathcal{F}^{-1}\{Z_1(\omega)I(\omega)\}.
\end{align}
\paragraph{Response to a zero-mean single-sine}
Let us now look at the response of an NLTI system, described by a Volterra series, to a zero-mean single-sine excitation $i(t)=I \cos(\omega t)$. The contribution of the $n$-th order nonlinearity to the voltage signal yields \cite{lang1996output}
\begin{subequations}
\begin{align}
v_n(t)=\sum_{h=0}^n \vert V_{n,h}\vert \cos\big(h\omega t +\angle V_{n,h}\big),
\label{eq:vnVolterraSeries}
\end{align}
with 
\begin{align}
V_{n,h}&=Z_{n,h}(\omega)I^n \qquad h>0\\
Z_{n,h}(\omega)&=\frac{1}{2^{n-1}} \sum_{\{s_1,\hdots,s_n\}\in\mathbb{S}_{n,h}} Z_n(s_1\omega,\hdots,s_n\omega). 
\end{align}
\label{eq:responseVolterraSeriesnSingleSine}
\end{subequations}

\noindent
For harmonic $h=0$, the premultiplying factor should be $1/2^n$ instead of $1/2^{n-1}$. The set $\mathbb{S}_{n,h}$ contains all possible lists $\{s_1,\hdots,s_n\}$, with elements $s_{1,\dots,n}\in \{-1,1\}$, such that $s_1+\hdots+s_n=h$. When the set is empty, $Z_{n,h}(\omega)=0$. The values $Z_{n,h}(\omega)$ are called the \emph{nonlinear impedance coefficients}, and come directly from the generalised impedances. A sum of $n$ elements with values that are either $1$ or $-1$ can never be larger than $n$ and the sum of an even number of these elements can never be odd, and vice versa. This translates into,
\begin{subequations}
\begin{align}
&Z_{n,h}(\omega)=0 & & \text{for } h>n\\
&Z_{2n,2h+1}(\omega)=0 & & \forall n,h \in \mathbb{N}\\
&Z_{2n+1,2h}(\omega)=0& & \forall n,h \in \mathbb{N}.
\end{align}
\label{eq:simplificationVnh}
\end{subequations}
\noindent These seemingly complicated mathematics are better understood by looking at a few special cases. For the linear term ($n=1$) in the Volterra series we find that $V_{1,0}=0$ and $\mathbb{S}_{1,1}=\{1\}$, such that $V_{1,1}=Z_1(\omega)I$. Hence, the voltage response yields
\begin{align}
    v_1(t)&=\underbrace{\vert Z_1(\omega)\vert I}_{\vert V_{1,1}\vert} \cos\big(\omega t+\underbrace{\angle Z_1(\omega)}_{\angle V_{1,1}}\big).
\end{align}
The response is present at the same frequency as the excitation, which is in accordance with the expected linear output \eqref{eq:responseSingleSine}. For nonlinear systems, that is $n_\mathrm{max}\geq2$, we notice from \eqref{eq:vnVolterraSeries} that spectral content may be present at \emph{integer} multiples of the excited frequency. Considering \eqref{eq:VolterraResponsen} and \eqref{eq:simplificationVnh} with purely quadratic ($n=2$) and purely cubic ($n=3$) nonlinearities, we find, respectively,
\begin{align}
    v_2(t)&=\vert V_{2,0}\vert +\vert V_{2,2} \vert \cos\big(2\omega t+\angle V_{2,2}\big)\nonumber\\
    v_3(t)&=\vert V_{3,1}\vert \cos\big(\omega t+\angle V_{3,1}\big)+\vert V_{3,3} \vert \cos\big(3\omega t+\angle V_{3,3}\big).
\end{align}
The expressions for the different $V_{x,y}$ terms are given in \ref{app:VolterraSeriesCoefficients}. We remark that quadratic nonlinearities introduce spectral content at the \emph{even} integer multiples of the excited frequency smaller or equal to two, and that cubic nonlinearities introduce spectral content at the \emph{odd} integer multiples of the excited frequency smaller or equal to three. These results can be generalised to higher order even and odd nonlinearities,
\begin{align}
    v_{2n}(t)&=\sum_{h=0}^n \vert V_{2n,2h}\vert \cos\big(2h\omega t+ \angle V_{2n,2h}\big)\nonumber\\
    v_{2n+1}(t)&=\sum_{h=0}^n \vert V_{2n+1,2h+1}\vert \cos\big((2h+1)\omega t+ \angle V_{2n+1,2h+1}\big).
\end{align}
Accordingly, the even and odd degree terms in the Volterra series introduce frequency content at, respectively, even and odd integer multiples of the excited frequency. Even nonlinear functions can be captured by the sum of even degree monomials of the Volterra series whereas odd nonlinear functions can be captured by odd degree monomials. Accordingly, even nonlinear behaviour is present at even integer multiples of the excited frequency, and odd nonlinear behaviour at odd multiples.

In contrast with the small excitation magnitude applied to the Butler-Volmer equation (Fig.~\ref{fig:nonlinearity1}), where the response is only present at the excited frequency $f_1$, for a larger excitation magnitude (Fig.~\ref{fig:nonlinearity2}), spectral content may also be present at integer multiples of the excited frequency. For this particular illustrative example, only odd nonlinear distortions are present, since we fixed $\alpha_a=\alpha_c=0.5$ and therefore the Butler-Volmer equation \eqref{eqn:BV} is an odd function around $0$. However, in practice, when $\alpha_a\neq\alpha_c$ the Butler-Volmer equation is neither even nor odd, and hence, both even and odd nonlinear distortions are present.

Writing out the analytical expression of the response of an NLTI system described by a Volterra series to a \emph{multisine} excitation is a more complicated matter. The mathematical details are omitted for this review paper, however, they are given in Lang and Billings \cite{lang1996output}. Fortunately, when the multisine consists of excited frequencies at integer multiples of a fundamental frequency, the observations made above are still valid. Nonlinearities will still be present at integer multiples of this fundamental frequency. However, the distinction between even and odd nonlinear distortions can only be made when \textit{only} odd integer multiples of the fundamental frequency are excited in the multisine. This is discussed in Section~\ref{Section:advantagesORPEIS}.

\subsubsection{Nonlinear EIS}\label{Section:NLEIS_theory}
It can be shown that the total response of the Volterra series of infinite order ($n_\mathrm{max}\rightarrow \infty$), excited by a sinusoidal signal at frequency $\omega$, $i(t)=I\cos(\omega t)$, introduces spectral content at all the integer multiples of the excited frequency $\omega$, that is,
\begin{subequations}
\begin{align}
    v(t)=\mathrm{OCV}+\sum_{h=0}^\infty \vert V_h\vert \cos(h\omega t+\angle V_h),
\end{align}
with 
\begin{align}
V_h&=\sum_{n=1}^\infty V_{n,h}=\sum_{n=h}^\infty Z_{n,h}(\omega)I^n\\
&=\sum_{r=0}^\infty Z_{h+2r,h}(\omega)I^{h+2r}.
\end{align}
\label{eq:VhNLEIS}
\end{subequations}

\noindent
In these equations, \eqref{eq:VolterraResponsen}, \eqref{eq:responseVolterraSeriesnSingleSine} and \eqref{eq:simplificationVnh} have been exploited. Only even order nonlinear impedance coefficients larger or equal to $h$ introduce spectral content at an even harmonic $h$, and vice versa for odd harmonics.

Nonlinear EIS \cite{wilson2006nonlinear,murbach2018nonlinear,kirk2022nonlinear} aims at measuring the \emph{leading order} nonlinear impedance coefficients $Z_{h,h}(\omega)$. These are defined from \eqref{eq:VhNLEIS} as \cite{kirk2022nonlinear},
\begin{align}
Z_{h,h}(\omega)=\lim_{I\rightarrow 0} \frac{V_h}{I^h}.
\label{eq:Zhhlim}
\end{align}
Note that the unit of the $h$-th leading order nonlinear coefficient $Z_{h,h}(\omega)$ is $\Omega/\text{A}^{h-1}$. In practice, the leading order nonlinear coefficients are measured as,
\begin{align}
\hat Z_{h,h}(\omega)= \frac{V_h}{I^h}=Z_{h,h}(\omega)+\underbrace{\sum_{r=1}^\infty Z_{h+2r,h}(\omega)I^{2r}}_\text{choose $I$ such that negligible},
\label{eq:Zhh}
\end{align}
where the excitation amplitude $I$ is chosen large enough such that the $h$-th harmonic $V_h$ is visible, but also small enough such that the higher order contributions $Z_{h+2r,h}(\omega)I^r$, $r\geq1$, are negligible. Note that under these measurement conditions, $Z_{1,1}(\omega)$ is the regular impedance $Z(\omega)$. The second leading order nonlinear impedance coefficient was measured for Li-ion batteries in \cite{murbach2018nonlinear} and \cite{kirk2022nonlinear}. Recall that the leading order nonlinear impedance coefficients are still models at a specific local operating point.

\subsubsection{The best linear approximation}
\begin{figure}
    \centering
    \includegraphics[width=0.5\textwidth]{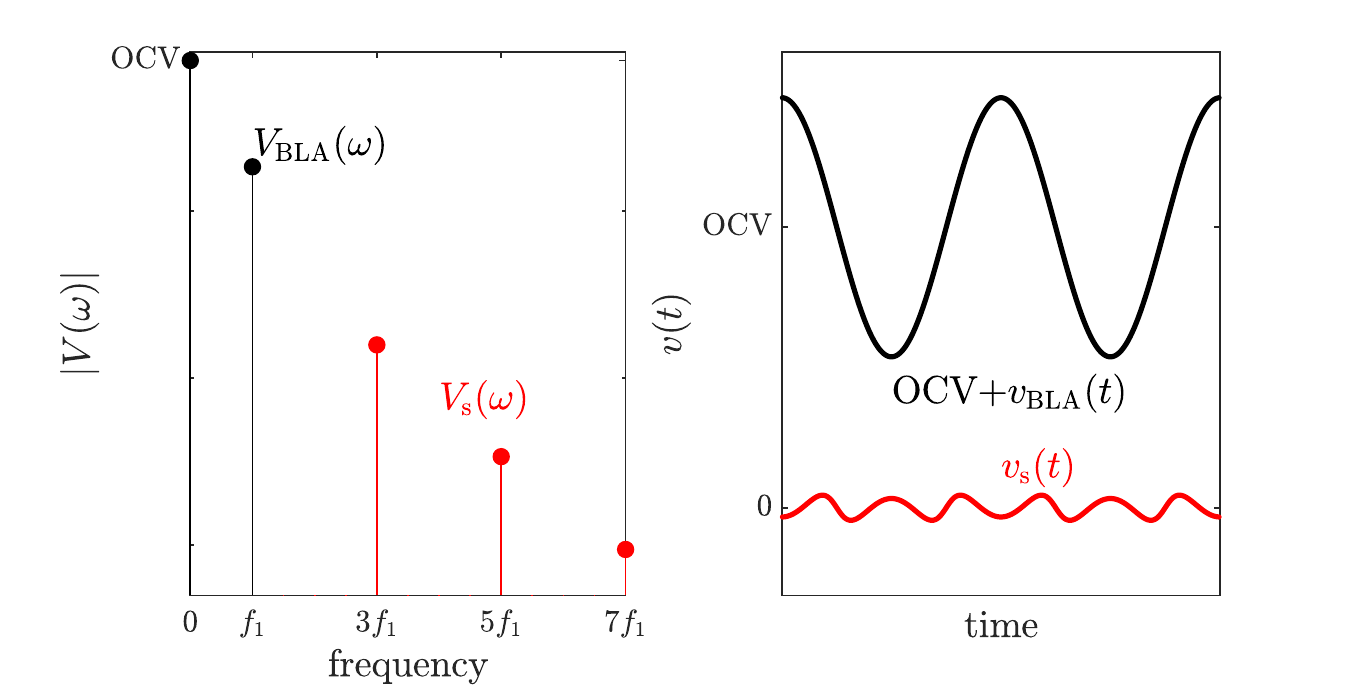}
    \caption{Decomposition of the nonlinear time-invariant response of Fig.~\ref{fig:nonlinearity2} into a linear response and purely nonlinear response. Left graph: frequency domain, right graph: time domain.}
    \label{fig:nonlinearity_lin_nonlin}
\end{figure}

For reasons of simplicity, we may prefer to work with linear models. In Schoukens et al.\ \cite{schoukens2005identification}, NLTI systems are modelled as so-called best linear approximations, plus a `nonlinear noise source' generating nonlinear distortions $v_\mathrm{s}(t)$\nomenclature[F]{$v_\mathrm{s}(t)$}{Nonlinear distortions}. Hence, in this context
\begin{align}
v(t)=\mathrm{OCV}+\underbrace{\mathcal{F}^{-1}\{Z(\omega)I(\omega)\}}_{v_\mathrm{BLA}(t)}+v_\mathrm{s}(t),
\end{align}
where $v_\mathrm{BLA}(t)$ stands for the response of the BLA $Z(\omega)$. The BLA is the `best' linear model in the sense that it minimises the nonlinear distortions in least square sense,
\begin{align}
Z=\arg \min_{Z'} \mathbb{E}\Biggl\{\int_{-\infty}^\infty \vert v_\mathrm{s}(t)\vert ^2 \mathrm{d}t\Biggr\},
\end{align}
with the expected value $\mathbb{E}\{\cdot\}$\nomenclature[O]{$\mathbb{E}\{\cdot\}$}{Expected value} taken over different realisation if the excitation signal \cite{enqvist2005linear}. Using Parceval's theorem, the BLA can also be defined in the frequency domain,
\begin{align}
Z(\omega)&=\arg \min_{Z'(\omega)} \mathbb{E}\Bigl\{|V(\omega)-Z'(\omega)I(\omega)|^2\Bigr\}.
\end{align}
It follows that the nonlinear distortions are uncorrelated with, but not independent of, the excitation signal. Hence, for a zero-mean single-sine excitation, the response of the BLA is located at the excited frequency, while the nonlinear distortions are present at the remaining integer multiples of the excited frequency,
\begin{subequations}
\begin{align}
v_\mathrm{BLA}(t)&=\vert V_{1}\vert \cos(\omega t +\angle V_{1})\\
v_\mathrm{s}(t)&=\vert V_0\vert+\sum_{h=2}^\infty \vert V_{h}\vert \cos(h\omega t +\angle V_{h}),
\end{align}
\end{subequations}
with the coefficients $V_h$ defined in \eqref{eq:VhNLEIS}. An illustration of this decomposition into a linear response and purely nonlinear response is depicted in Fig.~\ref{fig:nonlinearity_lin_nonlin}. The linear response is the one at the excited frequency, while the nonlinear response consists of the remaining frequencies. This BLA method has the major advantage that a linear model can still be justified when the nonlinear distortions are sufficiently small. Accordingly, it still makes sense to measure an impedance from data which shows nonlinear distortions, as long as these nonlinear distortions are small enough for the intended application. Therefore, it is important to detect and quantify the nonlinear distortions in measurements.

\begin{figure}
    \centering
    \includegraphics[width=0.5\textwidth]{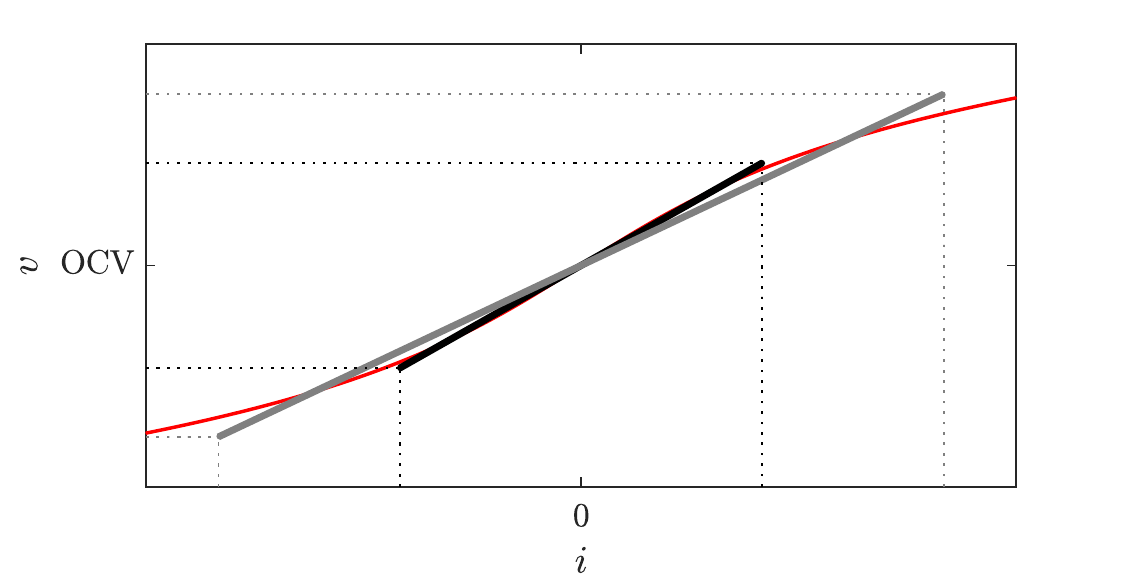}
    \caption{The linearisation of a static nonlinear function depends on the span it is linearised over. The red line represents the Butler-Volmer equation \eqref{eq:vZasFunctionOfiBV}. Linearising it over the black span gives a different BLA (the slope) than linearising it around the grey span.}
    \label{fig:BLAdependsonI}
\end{figure}

The BLA under a single-sine excitation is defined from \eqref{eq:VhNLEIS} as,
\begin{align}
Z(\omega)=\frac{V_1}{I}=\sum_{r=0}^\infty Z_{1+2r,1}(\omega)I^{2r},
\label{eq:DefinitionBLASingleSine}
\end{align}
and hence, \emph{the BLA depends on the amplitude of the excitation}. This is illustrated for the static case in Fig.~\ref{fig:BLAdependsonI}, the linearisation depends on the span over which the Butler-Volmer equation is linearised. The BLA here is the slope between voltage and current, which is clearly different for the black and grey lines. The reasoning in the dynamic case is similar.

Note that for nonlinear systems the first generalised impedance and the first leading order nonlinear impedance coefficients are equal, but the BLA impedance is different: $Z_1(\omega)=Z_{1,1}(\omega)\neq Z(\omega)$. This is because the higher order odd polynomials in the volterra series also contain a linear part, leading to a spectral contribution at the excited frequency that contributes to the BLA $Z(\omega)$.

\subsection{Nonstationary models}\label{Section:Nonstationarity}
The stationarity constraint in classical EIS experiments is very restrictive. The operating point, and hence also the external parameters, should be constant during an experiment. (NL)EIS can only be performed on systems in steady-state, resulting in models at fixed operating points. Accordingly, if we want impedance data over various operating conditions, for instance at different OCV values as in Fig.~\ref{fig:cEISOCV}, we have to separately drive the system to each of these operating conditions and wait for steady-state, which is very time-consuming. Moreover, sometimes the system never reaches steady-state due to inherently nonstationary behaviour caused by changes in thermodynamic states and kinetically slow side processes (such as the self discharge of energy storage systems). Furthermore, it is of great interest to study electrochemical systems during \emph{operation}. Examples include the formation of film layers during anodising, the electrorefining of copper, and the charging, discharging, and relaxation of batteries. For these examples, classical EIS or NLEIS can only be performed as a perturbation around a rest condition, while the evolution of the impedance during operation contains important information. Unfortunately, this information cannot be gathered by classical or stationary nonlinear EIS.

Nonstationarity can occur for two reasons, and they can happen simultaneously. The first cause is that the external parameters $p(t)$ vary during the experiment. The other cause is that the system is excited in such a way that it does not remain in steady-state during the experiment. This happens when superimposing a conventional excitation $i_\text{exc}(t)$ (see Section~\ref{Section:ExcitationSignalsClassicalEIS}) on a slow signal $i_0(t)$, driving the system in operating conditions,
\begin{align}
    i(t)=i_0(t)+i_\text{exc}(t).
\end{align}
For a battery, time-variation, for instance, occurs when the excitation is a multisine superimposed on a constant offset $i_0$ that (dis)charges the battery \cite{hallemans2022operando}, or when a zero-mean excitation is applied right after charging or discharging to study the relaxation behaviour \cite{koseoglou2021lithium}.

\subsubsection{The time-varying impedance}
Considering the two sources of nonstationarity, the linear voltage response can be modelled as,
\begin{align}
v(t)=v_0(t)+\int_{-\infty}^\infty z(\tau,t)i_\mathrm{exc}(\tau)\mathrm{d}\tau.
\label{eq:vtTVconv}
\end{align}
Here $v_0(t)$ represents a drift signal. In the battery example, this would be the voltage slowly increasing as the battery is charging due to a positive constant current. The time-variation due to the external parameters $p(t)$ and/or excitation trajectory $i_0(t)$ are simultaneously captured by a two-dimensional impulse response. This time-varying impulse response $z(\tau,t)$ was introduced by Zadeh \cite{zadeh1950frequency} in 1950 for modelling LTV systems. It is a natural extension of \eqref{eq:vconvTimeDomain}, where now the impulse response function explicitly depends on the excitation time.

It is also shown in Battistel et al.\ \cite{battistel2019physical} that nonstationarity \eqref{eq:vtTVconv} appears from an NLTI system when the response is linearised along a time-varying trajectory. This is detailed in \ref{app:linearisingOperatingTrajectory}. 

Similarly to \eqref{eq:vtZwIW_LTI}, the time-varying impedance $Z(\omega,t)$ appears when transforming \eqref{eq:vtTVconv} into the frequency domain \cite{zadeh1950frequency},
\begin{align}
v(t)=v_0(t)+\mathcal{F}^{-1}\{Z(\omega,t)I_\mathrm{exc}(\omega)\},
\label{eq:defTimeVaryingImpedance}
\end{align}
with the time-varying impedance defined as,
\begin{align}
Z(\omega,t)=\int_{-\infty}^\infty z(t-\tau,t)e^{-j\omega\tau}\mathrm{d}\tau.
\end{align}
\label{eq:DefinitionTimeVaryingImpedanceTotalab}
\noindent
Accordingly, when a single-sine excitation \eqref{eq:singleSine} is applied, superimposed on a slowly varying trajectory $i_0(t)$, the voltage response yields, 
\begin{align}
v(t)=v_0(t)+\vert Z(\omega_m,t)\vert I_m \cos\left(\omega_m t +\phi_m +\angle Z\left(\omega_m,t\right)\right).
\end{align}
The current excitation is, hence, modulated in amplitude and phase by $Z(\omega_m,t)$. Since linearity is still assumed, the response to a multisine excitation \eqref{eq:multisine1} is simply the sum of the responses to each separate sinusoidal signal,
\begin{multline}
v(t)=v_0(t)\:+\: \\ \sum_{m=1}^M\vert Z(\omega_m,t)\vert I_m \cos\left(\omega_m t +\phi_m +\angle Z\left(\omega_m,t\right)\right).
\end{multline} 
\begin{figure}
    \centering
    \includegraphics[width=0.5\textwidth]{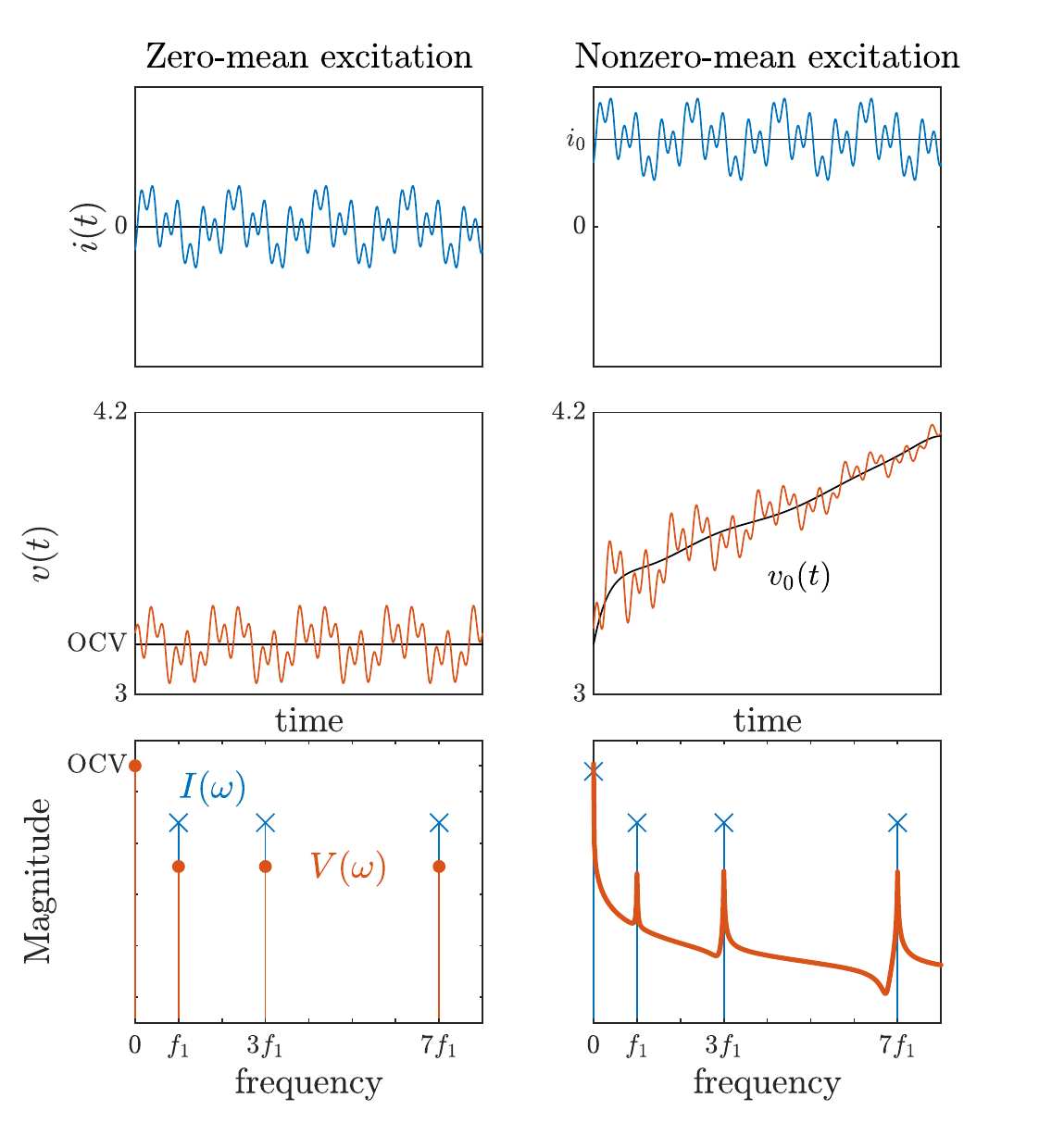}
    \caption{Illustration of the response of a Li-ion battery (which can be modelled by a Volterra series) to zero-mean and nonzero-mean excitations with small amplitudes. The slow parts $i_0(t)$ and $v_0(t)$ are shown in black. The positive mean value of the current charges the battery, and, hence, the voltage increases.}
    \label{fig:linearisationTrajectory}
\end{figure}
An illustration of the response of a Li-ion battery under zero-mean and nonzero-mean small amplitude excitation is shown in Fig.~\ref{fig:linearisationTrajectory}. For the zero-mean excitation, the response can be modelled as the OCV plus an LTI response. For the nonzero-mean excitation, nonstationarity is introduced due to the constant-current charging, which also might cause external parameters such as the temperature to change. Accordingly, the response can be modelled as a drift signal plus the response of an LTV system.

\subsubsection{Models along an operating trajectory} 
\begin{figure}
    \centering
    \includegraphics[width=0.5\textwidth]{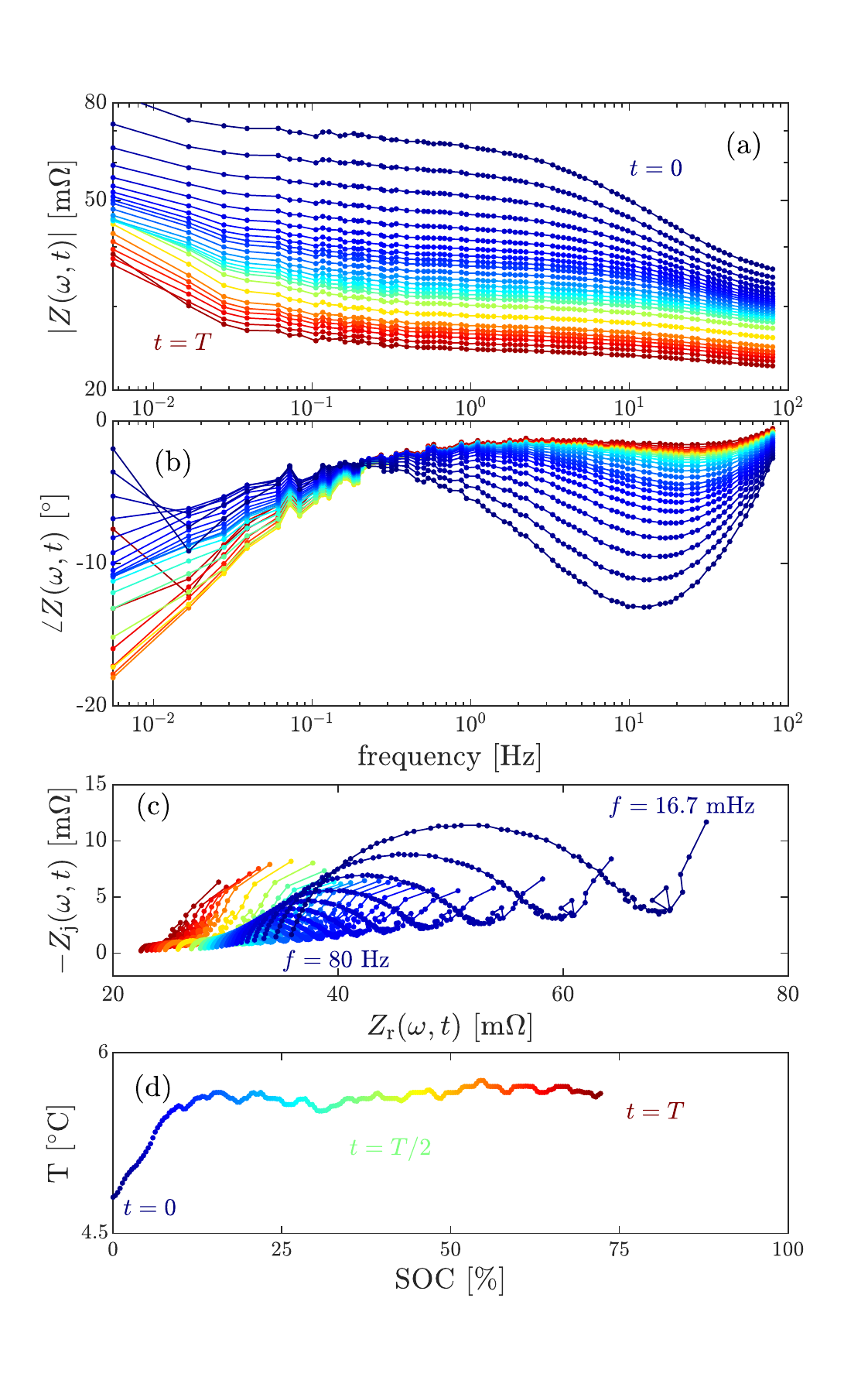}
    \caption{Illustration of time-varying impedance data of a Samsung 48X battery along an operating trajectory. The operating trajectory is caused by a charging current $i_0(t)=2.4$~A applied to a $4.8\,\mathrm{Ah}$ battery placed in a thermal chamber at $5^\circ$C. The time-varying impedance $Z(\omega,t)$ along the trajectory in the temperature-SOC plane (d) is shown as Bode plot (a-b) and Nyquist plot (c).}
    \label{fig:operatingTrajectoryImpedance}
\end{figure}

By choosing the slow trajectory $i_0(t)$ and/or varying external parameters during the experiment, we obtain a linear model for the impedance along an operating \emph{trajectory} instead of an operating \emph{point}. This trajectory is the drift of the system due to external effects during the measurement. One can, hence, obtain more global models than with classical EIS. This is illustrated in Fig.~\ref{fig:operatingTrajectoryImpedance}, where a battery with capacity $C$ = \SI{4.8}{Ah} is charged using a $C/2$ current, that is, $i_0(t)$ = \SI{2.4}{A}. Note that here the external temperature (a parameter), measured at the battery surface, changes due to thermal dynamics. The SOC also changes, however, this is not an external parameter, since it depends on the current excitation. 

\subsubsection{The importance of multisine excitation} \label{Section:theImportanceOfMultisineExcitations}
A multisine excitation is mandatory for accurate estimation of time-varying impedance. Since the system is changing over time during the experiment, it would be illogical to apply single-sines \emph{sequentially}, since then the impedance at each selected frequency would be computed for a different section of the operating trajectory. The advantage of using a multisine excitation is that many frequencies are excited \emph{simultaneously}, and we can obtain the impedance at all the selected frequencies over the \emph{entire} operating trajectory. In Section~\ref{Section:timeVaryingImpedanceEstimation}, we study how time-varying impedance data can be estimated from measured current and voltage data under multisine excitation.

\subsection{Nonlinear and nonstationary models}\label{section:nonlinearAndNonstationaryModels}
When a system is excited whilst not in steady-state, and with large excitation amplitudes $I_m$, nonstationary and nonlinear behaviour may happen \emph{simultaneously}. The system is then denoted as nonlinear time-varying (NLTV)\nomenclature[A]{NLTV}{Nonlinear time-varying}. In this case, the time-varying Volterra series with a superimposed drift signal provides a general model for the response,
\small
\begin{align}
v(t)=v_0(t)+\sum_{n=1}^{n_\mathrm{max}}\int_{-\infty}^\infty\cdots \int_{-\infty}^\infty z_n(\tau_1,...,\tau_n,t)\prod_{l=1}^n i_\mathrm{exc}(\tau_l)\mathrm{d}\tau_l.
\label{eq:VolterraResponsenltv}
\end{align}
\normalsize
Ideally, from such a time-varying Volterra series, we could measure time-varying leading order nonlinear impedance functions $Z_{h,h}(\omega,t)$. This would provide models over a large excitation amplitude and along a time-varying trajectory. However, this has, to the best of our knowledge, not been studied yet.

In Hallemans et al.\ \cite{hallemanstimevarying} we have studied the extension of the concept of BLA to BLTVA, that is, the best linear time-varying approximation. In this framework, the relation between current and voltage of an NLTV system is modelled as,
\begin{align}
v(t)=v_0(t)+\underbrace{\mathcal{F}^{-1}\{Z(\omega,t)I_\mathrm{exc}(\omega)\}}_{v_\mathrm{BLTVA}(t)}+v_\mathrm{s}(t),
\label{eq:defTimeVaryingImpandNL}
\end{align}
with $Z(\omega,t)$ the BLTVA and $v_\mathrm{s}(t)$ the time-varying nonlinear distortions. The BLTVA is a promising tool to monitor electrochemical system impedance during operation \cite{hallemans2022TrendRemoval,hallemans2022operando} (see Section~\ref{Section:operandoEIS}).

\section{Measuring current and voltage data}\label{Section:DataCollection}
In practical settings, we cannot measure continuous-time signals $i(t)$ and $v(t)$ over infinite periods. Instead, sampled current and voltage data over finite periods should be collected, denoted $i(n)$ and $v(n)$ where $n$ is the sample number. To obtain these, we apply an excitation through a potentiostat and measure sampled and windowed current and voltage data. For extracting time-varying impedance from this data, a multisine excitation is recommended, as used for odd random phase (ORP)\nomenclature[A]{ORP}{Odd random phase} EIS \cite{hallemans2022operando} and dynamic multi-frequency analysis \nomenclature[A]{DMFA}{Dynamic multi-frequency analysis} \cite{battistel2019physical}. 

\paragraph{Design of excitation signal}
Different kinds of excitation signals are `rich' enough to estimate classical impedance; among others, single-sines, multisines, and white noise. For obtaining stationary nonlinear impedance estimates, a single-sine excitation should be used. For obtaining time-varying impedance data, a multisine should be used. Since the single-sine excitation is a special case of the multisine, we focus on the latter.

A multisine with period $T_p$\nomenclature[C]{$T_p$}{Period length} superimposed on a time-varying trajectory $i_0(t)$ is given by 
\begin{align}
i(t)= i_0(t)+\underbrace{\sum_{m=1}^M I_m\cos\Big(\frac{2\pi h_m}{T_p} t+\phi_m\Big)}_{i_\mathrm{exc}(t)}.
\label{eq:multisine}
\end{align}
The trajectory $i_0(t)$ is user defined---for example this could be a charging or discharging current for a battery, or a chronoamperometry trajectory. Note that to obtain classical or stationary NLEIS data, this trajectory should be zero. The set of excited harmonics is defined as $\Hexc=\{h_1,h_2,...,h_M\}$\nomenclature[S]{$\Hexc$}{Excited harmonics}, and the excited angular frequencies are accordingly $\omega_{h_m}=2\pi h_m/T_p$, with $T_p$ the period of the multisine. The harmonic numbers should be integers, $h_m\in\mathbb{N}$, such that all sinusoidal signals fit an integer number of times in the period $T_p$. Note that the lowest frequency in the multisine ($f_1=1/T_p$) is inversely proportional to the period of the multisine. For a single-sine excitation, only $f_1$ is excited. In our definition, the natural numbers $\mathbb{N}$ do not include zero, while the set $\mathbb{N}_0$ does include zero. The amplitudes are selected by the user, depending on the application. The phases are most of the time chosen such as to minimise the crest factor of the overall multisine, that is, as not to introduce constructive interference when adding sines to each other. Different approaches can be used for this, including random phases picked from a uniform distribution in $[0,2\pi)$, and mathematical optimisation techniques such as the Schröder phase or DFT-based iterative algorithms to minimise the crest factor \cite{schoukens1988survey,schoukens1991design,guillaume1991crest,sanchez2011optimal,yang2015improved,zappen2018application}.

The name \emph{odd random phase} electrochemical impedance spectroscopy denotes impedance measurements under multisine excitation with only odd harmonics excited ($h_m\in2\mathbb{N}_0+1$), with random phases. This excitation signal was introduced by Hubin and Pintelon et al.\ \cite{van2009advantages,breugelmans2012odd}. In Section~\ref{Section:DetectionNLNS}, we show that inherent nonlinearity and nonstationarity in electrochemical systems can easily be detected using an ORP multisine excitation, and we discuss why it is advantageous to only excite \emph{odd} harmonics.

\paragraph{Windowing}
We cannot measure signals for an infinitely long time, but only for a certain period $t\in [0,T)$\nomenclature[C]{$T$}{Measurement time}. The measurement time $T$ is chosen to measure a certain ongoing reaction, for instance a charge cycle of a Li-ion battery. To avoid spectral leakage in the frequency domain, an \emph{integer} number of periods of a multisine excitation should be measured, that is, $T=PT_p$\nomenclature[C]{$P$}{Number of measured periods} with $P\in\mathbb{N}$ \cite[Section 2.2.3, p 40]{pintelon2012system}\footnote{Note that measuring an integer number of periods is only a requirement when the impedance estimation is performed in the frequency domain, which is the case for multisine experiments, but not necessarily for single-sine experiments.}. This is not always possible, but it is strongly recommended. Moreover, for obtaining NLEIS or time-varying impedance data, measuring an integer number of periods is a requirement.
\paragraph{Sampling}
Only a sampled representation of the continuous signal can be recorded, at a sampling frequency $f_s$\nomenclature[C]{$f_s$}{Sampling frequency}. Following the Shannon-Nyquist sampling theorem, this sampling frequency should be greater than twice the highest frequency in the measurements to avoid spectral aliasing. The sampling period is $T_s=1/f_s$\nomenclature[C]{$T_s$}{Sampling period}. It is important that the data is uniformly sampled.
\paragraph{Measuring the data} 
The sampled and windowed multisine current data is applied to an electrochemical device using a potentiostat. The potentiostat uses a digital-to-analog converter (DAC) to transform the generated time-series to a continuous signal. User-defined excitation is not always available in commercial potentiostats, but user-defined excitation is essential for the techniques in this article. Accordingly, multiple periods of the multisine excitation signal can be applied, and the potentiostat then measures the actual current and the voltage, which are also windowed and sampled. 

The collected data can be written as follows,
\begin{align}
\mathcal{D}_\text{time}=\left\{
    \begin{array}{ll}
         &[i(0),i(1),\hdots,i(N-1)]\\
         &[v(0),v(1),\hdots,v(N-1)]
    \end{array}
\right\},
\label{eq:sampledData}
\end{align}
where $x(n)$ is shorthand notation for $x(nT_s)$, $x=i,v$. The number of samples is $N=Tf_s$. 
\paragraph{frequency domain data}
Within the constraints of LTI systems, the impedance is defined as the ratio of the Fourier transforms of voltage and current \eqref{eq:ZomegaSimpleCont}. Hence, it would be appropriate to directly compute the impedance in the frequency domain. However, the Fourier transform acts on continuous signals, while only discrete-time measurements \eqref{eq:sampledData} can be collected from potentiostats. Fortunately, the spectrum of time-series can be computed by replacing the Fourier integral by a discrete sum. This is called the discrete Fourier transform\footnote{In 1965, Cooley and Tukey designed a highly-efficient algorithm to compute the DFT\nomenclature[A]{DFT}{Discrete Fourier transform}, which rapidly popularised frequency domain signal processing. This algorithm became known as the `fast Fourier transform' (FFT)\nomenclature[A]{FFT}{Fast Fourier transform} \cite{cooley1965algorithm} and is still used to date.} (DFT),
\begin{align}
X(k)=\frac{1}{N}\sum_{n=0}^{N-1}x(n)e^{-j2\pi kn/N}\quad k=0,1,...,N-1.
\end{align}
Here, $x(n)$ is again shorthand for $x(nT_s)$ and $x=i,v$. The DFT index $k$ corresponds to the angular frequency $\omega_k=2\pi k/T$ or frequency $f_k=k/T$. The frequency domain current and voltage data yields,
\begin{align}
\mathcal{D}_\text{freq}=\left\{
    \begin{array}{ll}
         &[I(0),I(1),\hdots,I(N-1)]\\
         &[V(0),V(1),\hdots,V(N-1)]
    \end{array}
\right\}.
\label{eq:sampledDataFreq}
\end{align}
When periodic time domain data is measured for an integer number of periods and sampled satisfying Shannon-Nyquist's theorem, the DFT coincides with the Fourier transform, evaluated on the DFT grid $f_k=k/T$ with $k=0,1,...,N-1$ \cite[Section 2.2, p 34]{pintelon2012system}.

\begin{figure}
    \centering
    \includegraphics[width=0.5\textwidth]{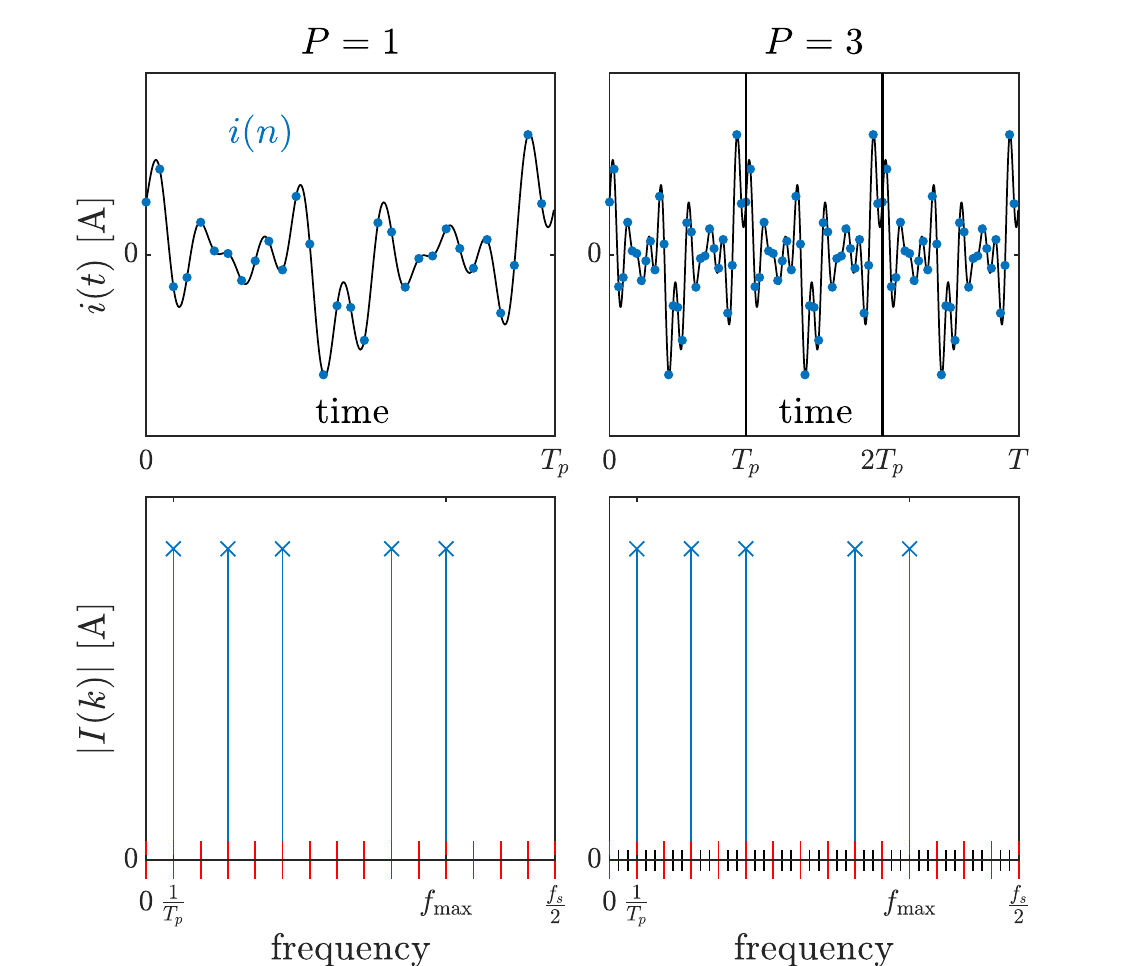}
    \caption{A sampled and windowed odd random-phase multisine in time and frequency domain. Top row: continuous signal (full line) and sampled data (blue dots). Bottom row: DFT of the sampled data, with the DFT grid indicated by red (original grid) and black (grid for $P=3$ measured periods) vertical bars on the frequency axis.}
    \label{fig:sampledMultisine}
\end{figure}

An illustration of one period of a windowed and sampled odd random phase multisine signal in the time and frequency domain, with the $7$-th harmonic unexcited for detection of odd nonlinear distortions, is shown in the left plots of Fig.~\ref{fig:sampledMultisine}. For the time domain plots, the full line is the continuous signal, while the dots are the sampled data. For the frequency domain plots, the DFT grid is indicated by vertical red bars on the frequency axis. Since DFT lines are intentionally left open at even harmonics and the odd harmonic $7$, it is possible to detect, quantify and classify nonlinear distortions in the response to one period of the excitation. However, if the system is also nonstationary, it is not possible to distinguish between nonlinear and nonstationary behaviour when measuring only one period.
\paragraph{Increasing the frequency resolution} Measuring more than one period increases the \emph{frequency resolution}, that is, the distance between two DFT lines $f_s/N=1/T$ becomes smaller. This is illustrated in the right plots of Fig.~\ref{fig:sampledMultisine}. For one measured period, the DFT grid is indicated in red, for three measured periods, in black. Nonlinearities in the spectrum of the response can \emph{only} be present at the DFT lines indicated in red, since these are the integer multiples of the fundamental excitation frequency $1/T_p$. The effect of nonstationarities can be present at all DFT lines, however. Accordingly, measuring a large number of periods is commonly used to distinguish between nonlinear and nonstationary behaviour, and the nonstationary behaviour can be modelled from the response spectrum.
\paragraph{Influence of measurement noise} We usually assume that noise in the measured data, $i_\text{meas}(n)$ and $v_\text{meas}(n)$, is additive, 
\begin{align}
x_\text{meas}(n)&=x(n)+\mathrm{n}_x(n)\qquad x=i,v,
\end{align}
and the noise time-series $\mathrm{n}_x(n)$ is iid (independent and identically distributed), zero-mean and Gaussian:
\begin{align}
\mathrm{n}_x(n)\sim \mathcal{N}(0,\sigma^2_{\mathrm{n}_x})\qquad x=i,v.
\end{align}
Here, $\sigma^2_{\mathrm{n}_i}$ and $\sigma^2_{\mathrm{n}_v}$ are the noise variances on the current and voltage, respectively. The DFTs of the noise time series, $\mathrm{N}_I(k)$ and $\mathrm{N}_V(k)$ are circular complex Gaussian\footnote{A two-dimensional Gaussian distribution on the complex plane.} distributed. However, the variances scale inversely with the number of samples $N$,
\begin{align}
\mathrm{N}_X(k)\sim\mathcal{N}_c\Big(0,\underbrace{\frac{\sigma^2_{\mathrm{n}_x}}{N}}_{\sigma^2_{\mathrm{N}_X}(k)}\Big)\qquad x=i,v.
\end{align}
We can interpret this result as follows: by measuring a higher number of samples $N$, the number of DFT lines increases, and, hence, the noise is distributed over more DFT lines, such that the variance of the circular complex distributed noise at each DFT line decreases. As a consequence, we can show that the frequency domain SNR\nomenclature[A]{SNR}{Signal-to-noise ratio} increases with the square root of the number of measured samples,
\begin{align}\label{eq:SNRforPeriodicExc}
\mathrm{SNR}_X(k)=\sqrt{\frac{\vert X(k)\vert^2}{\sigma^2_{\mathrm{N}_X}(k)}}=\sqrt{N}\frac{\vert X(k)\vert}{\sigma_{n_x}}.
\end{align}
Accordingly, by measuring a higher number of samples $N$\nomenclature[C]{$N$}{Number of measured samples}, we increase the SNR. In practice, the SNR is improved by increasing the number of measured periods $P$ of the data.
\section{Classical frequency domain impedance estimation}\label{Section:classicalImpedanceEstimationFD}
In the early seventies, Creason and Smith \cite{creason1972fourierI,creason1972fourierII,creason1973fourierIII} adopted the recently discovered FFT for classical EIS directly from frequency domain. These techniques, referred to as FFT-EIS, measure the impedance starting from DFT data \eqref{eq:sampledDataFreq}. An important question is the choice of the excitation signal. As explained earlier, a zero-mean excitation is needed for stationarity. In \cite{creason1973fourierIII}, different zero-mean excitation signals are studied: periodic excitations, transient inputs, and band-limited white noise. The conclusion is drawn that periodic excitation, with excited frequencies lying on the DFT grid, are superior. However, it is not always possible to apply periodic excitation and measure over an integer number of periods. Think for instance about a low-cost measurement apparatus for a BMS where only short and fixed-length data records can be handled. Hence, estimating techniques are also needed for random excitations. In 1975, Blanc et al.\ \cite{blanc1975etude} proposed the so-called `pseudo-white noise' technique (or `m\'ethode du bruit blanc', since this article was written in French). Pseudo-white noise may also be used as an excitation, and the impedance is computed as the ratio of the cross- and auto power spectra in the frequency domain. Howey et al.\  \cite{howey2013online} also used the ratio of the cross- and auto power spectra in the frequency domain to perform fast impedance measurements on a self-made low-cost excitation and measurement system for batteries that could be used in a BMS. The same issue of the choice of excitation signal for frequency domain system identification is studied in \cite{schoukens1988survey} and in Pintelon and Schoukens' book \cite[Chapter 2]{pintelon2012system}. We now mathematically formalise the impedance estimation for periodic and random excitation signals.

\begin{figure}
    \centering
    \includegraphics[width=0.5\textwidth]{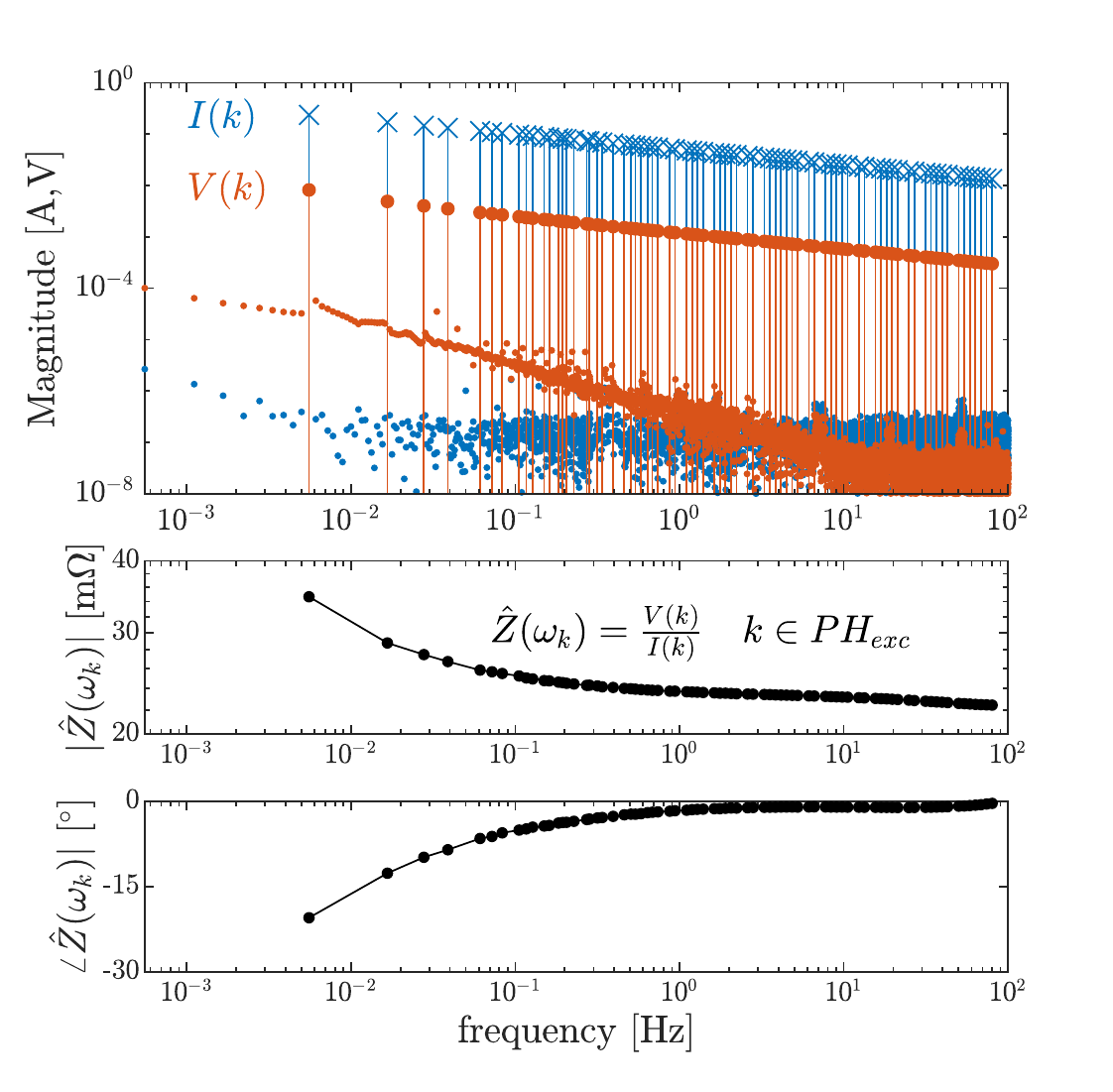}
    \caption{Example of classical impedance estimation for a Li-ion battery at $10$\% SOC and $25^\circ$C under a periodic multisine excitation measured for $P=10$ periods. The excited lines, $k\in P\Hexc$, are indicated with vertical lines with large dots and crosses. The remaining DFT lines (small dots) contain noise and a small drift signal.}
    \label{fig:MeasurementLiionIOspectrumLTI}
\end{figure}

\paragraph{Periodic excitation} When measuring an integer number of periods $P$ in steady-state under a periodic excitation, for instance a multisine, the DFT is exactly a sampled version of the continuous Fourier transform \cite[Section 2.2, p\ 34]{pintelon2012system}. Hence, the impedance can simply be computed from \eqref{eq:ZomegaSimpleCont} as the ratio of the voltage and current spectra,\nomenclature[O]{$\hat \cdot$}{Estimate}
\begin{align}
    \hat Z(\omega_k)=\frac{V(k)}{I(k)}& & \omega_k=\frac{2\pi k}{T},\ k\in P\Hexc,
    \label{eq:impedanceEstimateLTIPeriodicVoverI}
\end{align}
with $\Hexc$ the excited harmonics of the excitation signal. A measurement example is shown in Fig.~\ref{fig:MeasurementLiionIOspectrumLTI}. Since the OCV in the voltage is only present at DC\nomenclature[A]{DC}{Direct current - zero frequency} (zero frequency), it has no influence on the estimate of the impedance at positive frequencies. Note that in perfect LTI conditions, only measurement noise causes an uncertainty on the estimate \eqref{eq:impedanceEstimateLTIPeriodicVoverI}. Increasing the number of periods $P$ generates an averaging effect, which reduces this uncertainty \cite[Section 2.4, p 44]{pintelon2012system}. Periodic excitation is recommended when possible.
\paragraph{Random excitations} For random excitations, such as pseudo-white noise, pseudo-random binary sequences (PRBS) or multisine excitations not measured for an integer number of periods, the DFT does not correspond to the continuous Fourier transform anymore due to transients. These transients can be reduced by using windows, e.g.\ the Hann window. For random excitation, it is recommended to measure the impedance by the ratio of the cross- $\hat S_{VI}(k)$ and auto-spectra $\hat S_{II}(k)$ \cite[Section 2.6, p.\ 54]{pintelon2012system}, 
\begin{align}
    \hat Z(\omega_k)=\frac{\hat S_{VI}(k)}{\hat S_{II}(k)}=\frac{\sum_{m=1}^M V_{[m]}(k)I^*_{[m]}(k)}{\sum_{m=1}^M I_{[m]}(k)I^*_{[m]}(k)},
    \label{eq:impLTIRandomExcitation}
\end{align}
where the superscript $^*$ stands for the complex conjugate, and the subscript $_{[m]}$ stands for different experiments. Averaging over many different experiments reduces the transient and avoids divisions by zero. Note that this only makes sense in stationary conditions where multiple experiments can be gathered.

As a special case, Blanc \cite{blanc1975etude} chooses the pseudo-white noise excitation $i(t)$ such that $I(k)I^*(k)=1$ in the frequency band of interest. Accordingly, only the numerator of \eqref{eq:impLTIRandomExcitation} needs to be computed to estimate the impedance. Note that these cross- and auto spectra correspond to correlations in the time domain \cite{bendat1980engineering}, as also studied by Blanc \cite{blanc1975etude}.

Other, more involved, frequency domain techniques are possible for measuring the impedance using random excitation  \cite{lataire2016transfer,pintelon2020frequency}. Here, local parametric modelling and Gaussian process regression are used to separate the impedance from the transients. Such techniques are promising for low-cost experiments where periodic excitation signals cannot be applied. However, explaining these in detail goes beyond the scope of this article.

\section{Detection of nonlinearity and nonstationarity}
\label{Section:DetectionNLNS}
It is empirically shown in You et al.\ \cite{you2020application} that classical impedance data $\hat Z(\omega_k)$ obtained from broadband excitation using frequency domain techniques (as studied in Section~\ref{Section:classicalImpedanceEstimationFD}) always satisfies the Kramers-Kronig relations. Hence, the latter cannot assess to which extent measured multisine data satisfies the assumptions of linearity and stationarity, and another tool is required for this purpose. Inherently nonlinear and nonstationary behaviour of electrochemical systems is easily detected by applying a zero-mean ORP excitation \eqref{eq:multisine} (with $i_0(t)=0$) and studying the measured frequency domain data \eqref{eq:sampledDataFreq} \cite{d2005variance,van2009advantages,hallemans2020detection,hallemanstimevarying}. We recommend the use of this tool, which we now illustrate on a simplistic example that demonstrates its advantages.

\subsection{An example with ORP multisine excitation}
\begin{figure*}
    \centering
    \includegraphics[width=\textwidth]{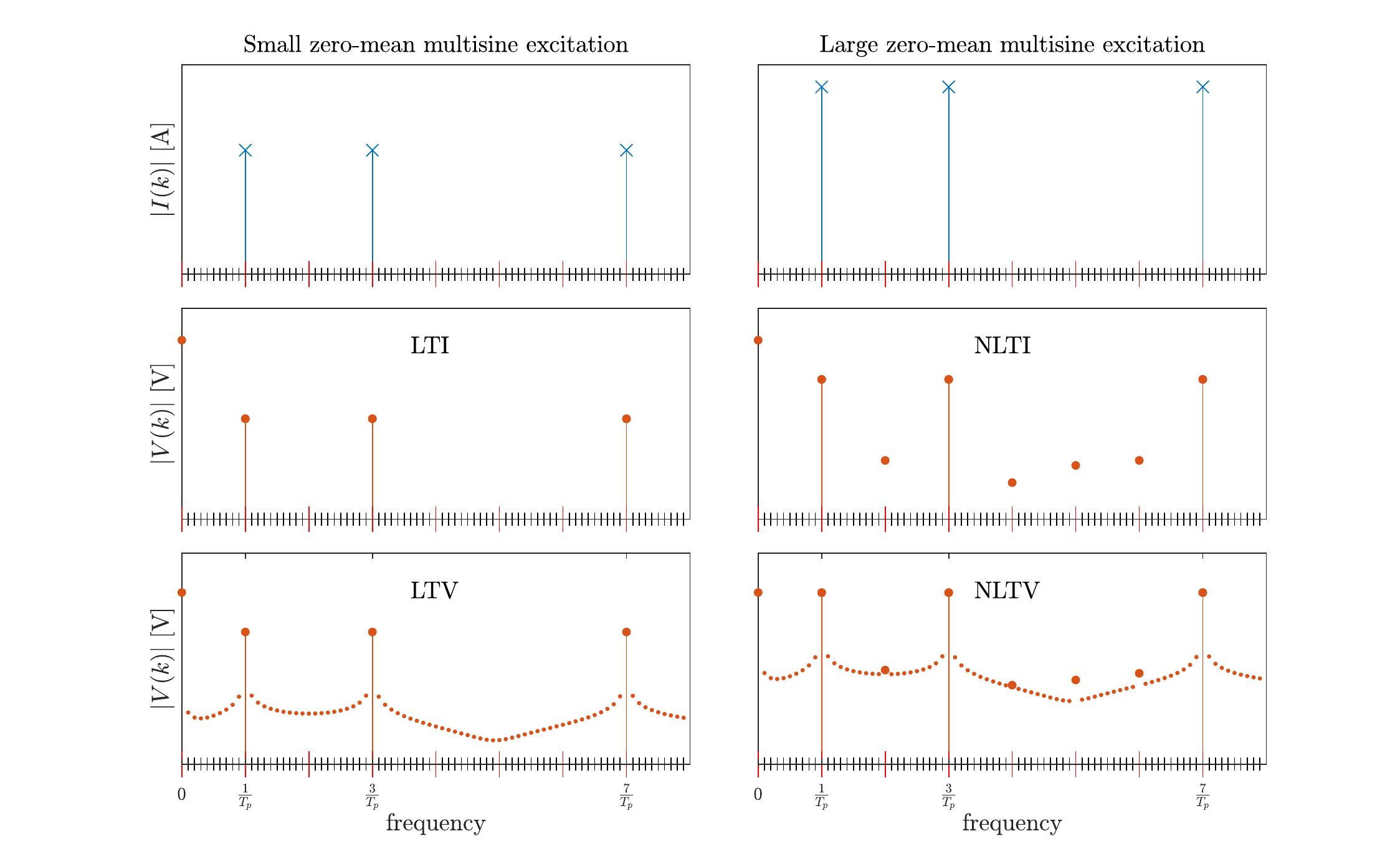}
    \caption{Detection of nonlinear and nonstationary behaviour in measured data. An odd random phase multisine signal (blue crosses) with a detector for odd nonlinear distortions is applied for $P=10$ periods at two different excitation amplitudes. In the LTI case, the voltage response only has spectral content at DC (dots at $f=0\,\mathrm{Hz}$) and at the excited frequencies. For a higher excitation amplitude, the system might become NLTI, and spectral content appears at integer multiples of the fundamental frequency. When the system is nonstationary, hyperbolic shapes become visible around the excited frequencies, and in the nonlinear case around all the integer multiples of the fundamental frequency.}
    \label{fig:spectra}
\end{figure*}

Consider an odd random phase multisine with excited harmonics $\mathbb{H}_\mathrm{exc}=\{1,3,7\}$, that is, excited frequencies $1/T_p$, $3/T_p$ and $7/T_p$. A measurement is performed for a duration $T=10T_p$, i.e., $P=10$ periods are measured. Based on the frequency domain voltage response data $V(k)$, it is possible to detect whether the electrochemical system behaves as an LTI, NLTI, LTV or NLTV system, as illustrated in Fig.~\ref{fig:spectra}.
\begin{description}
    \item [LTI:] $V(k)$ has only spectral content at DC and the excited harmonics $P\mathbb{H}_\mathrm{exc}$.
    \item [NLTI:] $V(k)$ has spectral content at DC and the excited frequencies, and also at harmonics that are integer multiples of the fundamental frequency $1/T_p$, i.e., $P\mathbb{H}_\mathrm{nl}$ with $\mathbb{H}_\mathrm{nl}=\{0,1,2,...\}$. Both even and odd nonlinearities are detected, since there is spectral content at the even left out harmonics ($0$, $2$, $4$, $6$) and at the odd left out harmonic ($5$).
    \item [LTV:] $V(k)$ consists of hyperbolic-like shapes around DC and the excited frequencies. These shapes, called \emph{skirts} in the literature \cite{lataire2012non}, are due to the smooth time-varying function modulating the multisine components.
    \item[NLTV:] $V(k)$ consists of hyperbolic-like shapes around DC, the excited frequencies and the nonlinear harmonics.
\end{description}
In Fig.~\ref{fig:spectra} noiseless data are considered. However, real-life measurements also contain noise that is distributed over all DFT frequencies. Still, it is easy to distinguish between nonlinearities, nonstationarities and noise. This can, for instance, be seen from Fig.~\ref{fig:MeasurementLiionIOspectrumLTI} where the measured voltage spectrum clearly satisfies the LTI constraints and the noise is at least $1000$ times smaller than the linear response. For a \emph{nonzero-mean} multisine excitation, the response will likely be nonstationary. This would look like a sampled version of Fig.~\ref{fig:linearisationTrajectory}. 

From all the spectra illustrated in Fig.~\ref{fig:spectra}, only the LTI one should be processed with classical EIS techniques, as discussed in the previous section. If only nonlinearities are detected, one can estimate the BLA (see Section~\ref{Section:BLAestimationfromdata}). If time-variation is detected, but no nonlinear distortions, time-varying impedance data can be estimated from the data record (see Section~\ref{Section:timeVaryingImpedanceEstimation}), if there are nonlinear distortions too, but they are limited, the BLTVA can be estimated from the data using operando EIS (see Section~\ref{Section:operandoEIS}).

\subsection{Advantages of an ORP multisine excitation} \label{Section:advantagesORPEIS}
By measuring the response of the electrochemical system to an ORP multisine excitation for a large number of periods, we can easily detect the presence of nonstationarities and nonlinearities, and estimate the noise level. It is possible to do this over a wide frequency range, in one single experiment, as studied in \cite{van2004electrochemical,van2006electrochemical,van2009advantages,hallemans2020detection}. The advantages of using a multisine excitation over a single-sine one have already been discussed in depth (Section~\ref{Section:theChoiceOfExcitationSignal}). However, an interested reader might wonder why it is particularly advantageous to use a multisine with \emph{random phases} and only \emph{odd} excited harmonics. The reasons are as follows:

The random phases are chosen to minimise the crest factor of the multisine. Note that there exist optimal ways to achieve this, however, using random phases has the advantage of simple implementation \cite{zappen2018application}.

The choice to excite only \emph{odd} harmonics, on the other hand, is more involved. In fact, the advantage is twofold. First, it allows distinguishing between even and odd nonlinear distortions. The nonlinear distortions are present at DFT lines which are the product of the excited harmonic numbers and the degrees of the relevant monomials in the Volterra series \cite{lang1996output}. When  exciting only odd harmonics, for odd nonlinear behaviour (odd degree monomials in the Volterra series), the nonlinear distortions are present at odd harmonics ($\mathrm{odd}\times\mathrm{odd}=\mathrm{odd}$), while for even nonlinear behaviour (even degree monomials in the Volterra series) they are present at even harmonics ($\mathrm{odd}\times\mathrm{even}=\mathrm{even}$). If we would only excite even harmonics, it would not be possible to distinguish between the two ($\mathrm{even}\times\mathrm{odd}=\mathrm{even}$ and $\mathrm{even}\times\mathrm{even}=\mathrm{even}$). When exciting all harmonics, even and odd, it is also not possible to distinguish between the types of nonlinearities.

The second advantage is that even nonlinearities do not have a contribution at the excited frequencies. Hence, when computing the BLA or BLTVA using a multisine excitation, even nonlinear distortions introduce no uncertainty, as further discussed in Section~\ref{Section:BLAestimationfromdata}. 

\section{Nonlinear impedance estimation}
\label{Section:NonlinearModelEstimation}
\subsection{Leading order nonlinear impedance estimation}
\label{Section:LeadingOrderNonlinearImpedanceEstimation}
It is appropriate to perform NLEIS (Section~\ref{Section:NLEIS_theory}) in the frequency domain. For this purpose, we apply a zero mean single-sine excitation, that is, $i_0(t)=0$, $M=1$ and $\Hexc=\{1\}$ in \eqref{eq:multisine}, measured over an integer number of periods $P\geq1$. When choosing the right amplitude $I_1$, as discussed in Section~\ref{Section:NLEIS_theory}, we obtain estimates of the leading order nonlinear impedance coefficients as,
\begin{align}
\hat Z_{h,h}(\omega_P)=\frac{V(hP)}{I(P)^h}& & \omega_P=\frac{2\pi P}{T}=\frac{2\pi}{T_p}.
\end{align}
Measuring a higher number of periods $P>1$ introduces an averaging effect of the stochastic noise in the experiments. Different frequencies $\omega_P$ are applied \emph{sequentially}. NLEIS with multisine excitation should also be possible, however not yet investigated. One has to be careful for leaving enough gaps in the excitation signal to see the integer harmonics in the response.
\subsection{Best linear approximation}\label{Section:BLAestimationfromdata}
Independently of the excitation amplitude of \emph{single-sine} experiments, the BLA can be estimated from \eqref{eq:DefinitionBLASingleSine}, where different frequencies can be applied \emph{sequentially},
\begin{align}
\hat Z(\omega_P)=\frac{V(P)}{I(P)} & & \omega_P=\frac{2\pi P}{T}=\frac{2\pi}{T_p}.
\end{align}
For \emph{multisine} excitations measured for an integer number of periods $P$, the BLA impedance estimate is computed exactly as for classical impedance experiments \eqref{eq:impedanceEstimateLTIPeriodicVoverI}. If an ORP multisine is used, that is, exciting odd harmonics, only odd nonlinearities will introduce spectral content at the excited frequencies. Indeed, even nonlinear distortions can only be present at even multiples of the odd harmonics, resulting in even harmonics. Accordingly, the BLA will have uncertainties due to noise and also due to odd nonlinear distortions that introduce spectral content at other odd harmonics (see \cite[Section 3.4, p. 78]{pintelon2012system}). The effect of the noise can again be reduced by measuring a larger number of periods. The odd nonlinear distortions, on the other hand, do not depend on the number of measured periods. They only depend on the system and the excitation, that is, the amplitudes and excited frequencies. The uncertainty of the BLA can be reduced by exciting fewer frequencies in the multisine. Recall that the BLA depends on the excitation \eqref{eq:DefinitionBLASingleSine}, however, using Riemann-equivalent multisines, one can adapt the number of excited frequencies without changing the BLA \cite{schoukens2009robustness}.

\section{Time-varying impedance estimation}
\label{Section:timeVaryingImpedanceEstimation}
Estimating the time-varying impedance $Z(\omega_k,t)$ cannot simply be done by dividing DFT spectra. Different approaches have been developed over the last two decades to reveal the time-varying impedance from data. Both single-sine and multisine techniques exist. However, in this review paper we intentionally restrict to multisine excitations, since we believe this is the only correct solution (see Section~\ref{Section:theImportanceOfMultisineExcitations}). With ever-growing computation power, more and more complex techniques have been developed. There are two main approaches. One implies the use of windowing/filtering of the current and voltage data in time or frequency domain, respectively, and gives an average value of the impedance inside the selected time frame. Another uses the mathematical regression of the voltage response spectrum to extract the time-variation. These approaches are detailed next, chronologically.

\subsection{FFT-EIS applied to nonstationary data}
The first attempt to obtain time-varying impedance data was by Bond, Schwall and Smith in 1977 \cite{bond1977line,schwall1977high,schwall1977line}. This simple and intuitive approach is an extension of Smith's FFT-EIS discussed in Section~\ref{Section:classicalImpedanceEstimationFD}.

Instead of applying only a zero-mean excitation signal, the excitation signal is now superimposed on a slower cyclic voltammetry excitation, introducing nonstationarity in the measured system. While collecting current and voltage data, the FFT-EIS is performed on short subrecords. This is a type of windowing. Accordingly, for each subrecord, which corresponds to a certain point in the voltage trajectory of the cyclic voltammetry, the time-averaged impedance of the subrecord is computed using the techniques of Section~\ref{Section:classicalImpedanceEstimationFD}. Basically, classical impedance measurements are applied to nonstationary data, but in very small time-windows such that the impedance could be assumed constant within each subrecord.

Since the period length is inversely proportional to the frequency, the impedance is only measured at high frequencies, such that short subrecords can be taken. Results of the impedance (or admittance) varying over the cyclic voltammetry trajectory are only shown for two particular frequencies above $300\,\mathrm{Hz}$. This is a strong limitation of the technique.

Later, in 1997, FFT-EIS was implemented on a micro computer \cite{hazi1997microcomputer}, for a multisine superimposed onto a staircase DC ramped voltage excitation. Time-varying impedance measurements could be obtained within the more broadband frequency range of $50\,\mathrm{Hz}$ -- $50\,\mathrm{kHz}$. Later, Sacci and Harrington \cite{sacci2009dynamic,sacci2014dynamic}, developed measurement apparatus to obtain time-varying impedance data using FFT-EIS with multisine excitation superimposed on a cyclic voltammogram.

\paragraph{Comment} It is noteworthy that obtaining time-varying impedance data is much easier at high frequencies. This on the one hand due to low frequency noise (so-called $1/f$ noise), but also due to the drift signal (or also called trend) $v_0(t)$ in \eqref{eq:defTimeVaryingImpedance} having a decreasing shape over frequency, and hence, hiding low-frequency content \cite{hallemans2022TrendRemoval,hallemans2022operando}. Moreover, for logarithmically distributed excited frequencies, nonstationarities are more easily detected at the high frequencies since the excited frequencies are more separated (expressed in DFT lines), thus making the `skirts' better visible. Remarkably, time-varying behaviour is mostly present at the low frequencies, this is possibly due to mass transport and/or charge transfer kinetics. Also, when measuring at high frequencies, the measurement time is usually shorter, therefore nonstationarities are smaller.

\subsection{Time-frequency analysis methods}
During the nineties, time-frequency analysis, as described by L. Cohen\footnote{Not to confuse with the great Canadian singer-songwriter.} \cite{CohenTimeFrequency}, became a widely used tool in signal processing. Time-frequency analysis describes how the spectral content of a signal $x(t)$ is changing in time, which is exactly needed for time-varying impedance estimation. The workhorse for this job is the short-time Fourier transform (STFT)\nomenclature[A]{STFT}{Short-time Fourier transform}, which computes the Fourier transform of a signal restricted by a window function $w(t)$,
\begin{align}
\mathrm{STFT}\{x\}(\omega,t)&=\int_{-\infty}^\infty w(t'-t)x(t')e^{-j\omega t'}\mathrm{d}t' \nonumber \\
&=\mathcal{F}\{w(t'-t)x(t')\},
\end{align}
with the Fourier transform acting on the variable $t'$. The most commonly used window functions are the Gaussian, Hamming and Blackmann-Harris windows. These windows reach their largest values in the center, and decrease smoothly towards the borders.
\subsubsection{STFT-EIS} 
After the millenium change, Darowicki \cite{darowicki2000theoretical,darowicki2000instantaneous,darowicki2003dynamic} proposed to estimate the time-varying impedance under multisine excitation as the ratio of the STFT of voltage and current,
\begin{align}
Z(\omega,t)=\frac{\mathcal{F}\{w(t'-t)v(t')\}}{\mathcal{F}\{w(t'-t)i(t')\}},
\label{eq:impedanceDarowicki}
\end{align}
with again the Fourier transforms acting on the variable $t'$. Assuming the window $w(t)$ to be a symmetric function centered around zero, the impedance at time $t$ is computed by selecting the time domain data around this time-instant, and dividing the corresponding spectra of voltage and current. Note that FFT-EIS on nonstationary data is a special case of STFT-EIS, with a rectangular window $w(t)$.

In practice, of course, only discrete-time data \eqref{eq:sampledData} is available. The impedance can then be computed by \cite{darowicki2003dynamic},
\begin{subequations}
\begin{align}
\hat Z(\omega_k,t_{n})=\frac{V(k,n)I^*(k,n)}{I(k,n)I^*(k,n)}& &  t_n=nT_s,
\label{eq:TVimpDarowickiDT}
\end{align}
with the DFT acting on subrecords of $N_w$ data points centered around $n$, which are windowed by the function $w(t)$,
\begin{align}
X(k,n)&=\frac{1}{N_w}\sum_{n'=n-N_w/2}^{n+N_w/2-1}w(n'-n)x(n')e^{-j2\pi kn'/N_w},
\end{align}
\end{subequations}
$x=i,v$. A division of the cross- and auto spectra is chosen here since the signals are not periodic anymore, and this estimator is recommended for random excitations (see Section~\ref{Section:classicalImpedanceEstimationFD}). The time-varying impedance estimate is computed at the harmonics $k$ where the numerator of \eqref{eq:TVimpDarowickiDT} shows peak values, corresponding to the excited frequencies of the multisine. To make these peaks visible and not overlapping too much, it is important that enough periods are measured. Moreover, the choice of the window $w(t)$ is crucial in this technique. 

\subsubsection{Dynamic Multi-Frequency Analysis} 
Later, Battistel and La Mantia \cite{koster2017dynamic,battistel2019physical,pianta2022evaluation} proposed the dynamic multi-frequency analysis (DMFA) for estimating time-varying impedance data. Here, the time-varying impedance is computed by filtering the current and voltage spectra around the excited frequencies of a multisine, and taking the inverse Fourier transform,
\begin{align}
Z(\omega,t)=\frac{\mathcal{F}^{-1}\{W(\omega'-\omega)V(\omega')\}}{\mathcal{F}^{-1}\{W(\omega'-\omega)I(\omega')\}},
\label{eq:impedanceLaMantia}
\end{align}
with the inverse Fourier transforms acting on $\omega'$. The function $W(\omega)$ implements a filtering operation. This process is called \emph{quadrature filtering} since only the spectrum at positive frequencies is considered and the inverse Fourier transforms give complex-valued results.

For sampled frequency domain data \eqref{eq:sampledDataFreq}, with multisine excitation, the time-varying impedance data estimates translate into 
\begin{subequations}
\begin{align}
\hat Z(\omega_k,t_n)=\frac{v(k,n)}{i(k,n)},
\end{align}
with for $k\in P\Hexc$ the inverse DFT acting on frequency domain subrecords of $N_W$ data points centered around $k$, which are filtered by the function $W(f)$,
\begin{align}
x(k,n)=\sum_{k'=k-N_W/2}^{k+N_W/2-1}W(k'-k)X(k')e^{j2\pi k'n/N_W} \quad  x=i,v.
\label{eq:iDFTwindowed}
\end{align}
\end{subequations}
Here also, it is important that enough periods are measured, such that around the excited frequencies, enough frequency domain data is available to extract the time-variation. 
\subsubsection{Equivalence between STFT-EIS and DMFA}
For a symmetrical window, $w(t)=w(-t)$, and $W(\omega)=\mathcal{F}\{w(t)\}$, the continuous definitions of the STFT-EIS \eqref{eq:impedanceDarowicki} and the DMFA \eqref{eq:impedanceLaMantia} are equivalent (as proven in \ref{app:Equivalence}),
\begin{align}
\frac{\mathcal{F}\{w(t'-t)v(t')\}}{\mathcal{F}\{w(t'-t)i(t')\}}=\frac{\mathcal{F}^{-1}\{W(\omega'-\omega)V(\omega')\}}{\mathcal{F}^{-1}\{W(\omega'-\omega)I(\omega')\}}.
\label{eq:STFTDMFAequivalenceContinuous}
\end{align}
Since the Fourier and inverse Fourier transforms act on, respectively, $t'$ and $\omega'$, both left and right hand side of \eqref{eq:STFTDMFAequivalenceContinuous} are a function of $\omega$ and $t$. Note that both latter equations for extracting the time-varying impedance are heuristic, and do not exactly match with the theoretical definition \eqref{eq:DefinitionTimeVaryingImpedanceTotalab}. Nonetheless, these approximations may be accurate enough in practice.

Since the STFT-EIS and DMFA approaches have a mathematically equivalent definition of the impedance, the difference boils down to the choice of the window $w(t)$, or equivalently the filter $W(\omega)$, and the actual implementation. The properties of the symmetrical window or filter can mainly be studied by its width. This width can, for instance, be defined by the variances
\begin{align}
\sigma^2_t= \frac{\int_{-\infty}^\infty t^2 \vert w(t)\vert^2 \mathrm{d}t}{\int_{-\infty}^\infty \vert w(t)\vert^2 \mathrm{d}t}\ \text{and} \ \sigma^2_\omega= \frac{\int_{-\infty}^\infty \omega^2 \vert W(\omega)\vert^2 \mathrm{d}\omega}{\int_{-\infty}^\infty \vert W(\omega)\vert^2 \mathrm{d}\omega}.
\end{align} 
Similarly as in quantum mechanics where the uncertainty principle prohibits to measure simultaneously the position and velocity of an electron with arbitrary precision, also in this case it is not possible to measure the impedance with arbitrary precision in both time and frequency. Accordingly, the so-called \emph{Gabor limit} \cite{CohenTimeFrequency} states that
\begin{align}
\sigma^2_t \sigma^2_\omega\geq\frac{1}{4}.
\end{align}
The time-selectivity $\sigma^2_t$ and frequency resolution $\sigma^2_\omega$ cannot both be made arbitrarily small. One has to trade-off between one and the other.

\paragraph{STFT-EIS} Darowicki \cite{darowicki2000theoretical} uses a Gaussian window, which has the property that its Fourier transform is Gaussian as well,
\begin{align}
w(t)=e^{-\frac{\lambda}{2}t^2}\  \iff \  W(\omega)=\sqrt{\frac{2\pi}{\lambda}}e^{-\frac{\omega^2}{2\lambda}}.
\end{align}
For this choice of window, the Gabor limit reduces to
\begin{align}
\sigma^2_t \sigma^2_\omega=\frac{1}{4} \quad \text{with}\quad \lambda=\frac{1}{2\sigma^2_t}=2\sigma^2_\omega.
\end{align}
Accordingly, an increase of the time selectivity leads to a deterioration of the frequency resolution, and vice versa. The time and frequency resolution are determined by the hyper parameter $\lambda$. The larger $\lambda$, the more resolution in time, but the less resolution in frequency, resulting in a trade-off.

An issue with STFT-EIS is that first subrecords are taken, followed by the windowing. As a consequence, we do not actually reach the Gabor limit. Also the DFT is taken on subrecords, which obviously contain less data points than the total record, hence, as discussed in Section~\ref{Section:DataCollection}, the SNR is poorer. Another issue is obtaining time-varying impedance data at low frequencies. For being able to measure low frequencies, the width of the window should, at least, comprise a period of this frequency. The period length is inversely proportional to the frequency, hence, for measuring low frequencies, the window width should be large, deteriorating the time resolution. On the other hand, since STFT-EIS is applied to time windows, the advantage is that the time-varying impedance estimation can be done in real-time.

\paragraph{DMFA} Battistel and La Mantia \cite{battistel2019physical}, on the other hand, directly define the quadrature filter,
\begin{align}
W(\omega)=\frac{\big(1+e^{-q^2}\big)^2}{\big(1+e^{-q\frac{\omega+\Delta \omega}{\Delta \omega}}\big)\big(1+e^{q\frac{\omega-\Delta \omega}{\Delta \omega}}\big)},
\end{align}
where $q$ is a factor determining the roll-off of the filter and $\Delta \omega$ is its bandwidth. Note that in the limits, we have that
\begin{subequations}
\begin{align}
\lim_{q\rightarrow 0} W(\omega)=1
\end{align}
and
\begin{align}
\lim_{q\rightarrow \infty} W(\omega)=\left\{
    \begin{array}{ll}
        1 & \mbox{if } -\Delta\omega\leq \omega\leq \Delta \omega \\
        0 & \mbox{else.}
    \end{array}
\right.
\end{align}
\end{subequations}
The objective of this filter is to mimic a rectangular filter, while being continuous.

The advantage of the DMFA over the STFT-EIS is that only one DFT should be performed on the entire time-series data record \eqref{eq:sampledData} to obtain the frequency domain data \eqref{eq:sampledDataFreq}. Here, the measurement noise is distributed over all the DFT lines, resulting in a higher SNR (see Section~\ref{Section:DataCollection}). The time-varying impedance data is then directly obtained by applying the inverse DFT to small windowed subrecords around the excited frequencies of the multisine. This has advantages in computation time (see \cite{battistel2019physical} Section 2.3). Also the width of the filter can be chosen in function of the spacing of the excited frequencies, possibly different for each excited frequency. Moreover, an analysis of the influence of measurement noise on the time-varying impedance data was performed \cite{chukwu2022statistical}. However, the time-varying impedance estimation cannot be done in real-time.

The main advantage of these time-frequency analysis methods is the accessible implementation. However, both time-frequency analysis methods have difficulty in estimating time-varying impedance data at low frequencies, this is due to the drift signal. Moreover, they do not account for treating nonlinear distortions, even though in the DMFA this could be implemented. These problems are each solved by operando EIS, as detailed next.

\subsection{Operando EIS}\label{Section:operandoEIS}
Operando EIS, as developed by Hallemans et al.\ \cite{hallemanstimevarying,hallemans2022TrendRemoval,hallemans2022operando}, is an extension of ORP-EIS. Here, we estimate the time-varying impedance by using the definition of the BLTVA \eqref{eq:defTimeVaryingImpandNL} as a model structure. Note that if no nonlinear distortions are present in the measurements, the BLTVA and the time-varying impedance in \eqref{eq:DefinitionTimeVaryingImpedanceTotalab} are equal. Nonlinearities in the measurements are detected, quantified and classified, and the noise level is estimated \cite{hallemanstimevarying}. Also uncertainty bounds are included on the estimated impedance data. Moreover, drift signals are suppressed, allowing to access low frequencies \cite{hallemans2022TrendRemoval}. 

The idea is to write the time-varying impedance as a truncated series expansion in a set of known basis functions in time,
\begin{align}
Z(\omega,t)=\sum_{p=0}^{N_p}Z_p(\omega)b_p(t).
\label{eq:timeVaryingImpAssumptionOperandoEIS}
\end{align}
\begin{figure}
    \centering
    \includegraphics[width=0.5\textwidth]{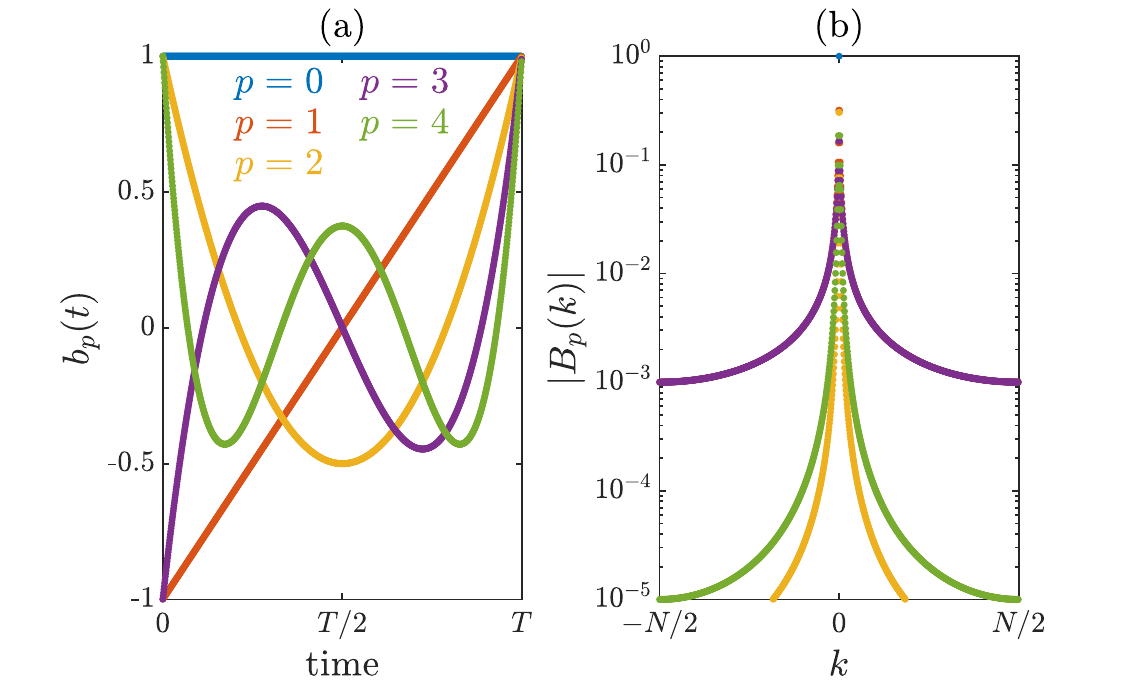}
    \caption{The first four Legendre polynomials in time (a) and frequency domain (b).}
    \label{fig:bpBp}
\end{figure}

\noindent
The basis functions $b_p(t)$\nomenclature[F]{$b_p(t)$}{Legendre polynomial basis functions} are chosen as Legendre polynomials (\ref{app:LegPol} and Fig.~\ref{fig:bpBp} (a)), since these benefit from good numerical conditioning \cite{lataire2012non}. Using \eqref{eq:timeVaryingImpAssumptionOperandoEIS}, the frequency and time dependencies are separated. Since the basis functions are known, only the impedances $Z_p(\omega)$ should be estimated. 

The time-varying nonlinear distortions $v_\mathrm{s}(t)$ are also expanded in series,
\begin{align}
v_\mathrm{s}(t)=\sum_{p=0}^{N_p}v_{\mathrm{s},p}(t)b_p(t),
\label{eq:timeVaryingNLbp}
\end{align}
with $v_{\mathrm{s},p}(t)$ the time-invariant nonlinear distortions generated by a Volterra series \eqref{eq:VolterraResponsen}, meaning that $V_{\mathrm{s},p}(\omega)=\mathcal{F}\{v_{\mathrm{s},p}(t)\}$ is only nonzero at the integer multiples of the fundamental frequency of the periodic excitation. Plugging \eqref{eq:timeVaryingImpAssumptionOperandoEIS} and \eqref{eq:timeVaryingNLbp} in the excitation-response relation of NLTV systems \eqref{eq:defTimeVaryingImpandNL} yields,
\begin{align}
v(t)=v_0(t)+\sum_{p=0}^{N_p}\mathcal{F}^{-1}\{Z_p(\omega)I_\mathrm{exc}(\omega)+V_{\mathrm{s},p}(\omega)\}b_p(t).
\label{eq:vtZpNLTVcontinuous}
\end{align}
The drift signal $v_0(t)$ is unknown and hides low frequency content. Therefore, it should also be modelled, for instance by Legendre polynomials \cite{hallemans2020detection,hallemanstimevarying},
\begin{align}
v_0(t)=\sum_{q=0}^{N_q} \theta_q b_q(t).
\end{align}
Drift signals can also be removed by differencing, as detailed in \cite{hallemans2022TrendRemoval}. This has better performance, however, the mathematics are more involved, so for this review paper we restrict to modelling the drift signal with basis functions. 

Taking the DFT of \eqref{eq:vtZpNLTVcontinuous} gives,
\begin{align}
V(k)=V_0(k)+\sum_{p=0}^{N_p}\Big(Z_p(\omega_k)I_\mathrm{exc}(k)+V_{\mathrm{s},p}(k)\Big)\ast B_p(k),
\label{eq:VkNLTVBp}
\end{align}
with 
\begin{align}
V_0(k)=\sum_{q=0}^{N_q}\theta_qB_q(k).
\end{align}
Here it was used that a product in the time domain becomes a convolution in the frequency domain and $B_p(k)$ is the DFT of the Legendre polynomials, shown in Fig.~\ref{fig:bpBp} (b). For a multisine excitation measured over an integer number of periods $P$, we have that $I_\mathrm{exc}(k)$ is only nonzero at the harmonics $P\mathbb{H}_\mathrm{exc}$ and $V_{\mathrm{s},p}(k)$ is only nonzero at the harmonics $P\mathbb{H}_\mathrm{nl}$, with $\mathbb{H}_\mathrm{nl}=\{0,1,2,3,...\}$\nomenclature[S]{$\mathbb{H}_\mathrm{nl}$}{Nonlinear harmonics}. Note that $\mathbb{H}_\mathrm{exc}\subset \mathbb{H}_\mathrm{nl}$ . Accordingly, \eqref{eq:VkNLTVBp} can be simplified as,
\begin{align}
V(k)=\sum_{q=0}^{N_q}\theta_qB_q(k)+\sum_{p=0}^{N_p}\sum_{k'\in P\mathbb{H}_\mathrm{nl}} \theta_p(k')B_p(k-k'),
\label{eq:VkNLTVBpwithoutConv}
\end{align}
with 
\begin{align}
\theta_p(k')=Z_p(\omega_{k'})I_\mathrm{exc}(k')+V_{\mathrm{s},p}(k').
\end{align}
Eq.~\eqref{eq:VkNLTVBpwithoutConv} is linear in the parameters $\theta$, hence, the data can be written in matrix form,
\begin{align}
V=K\theta+N_v,
\end{align}
where $V$ is a stacked vector of the measured voltage spectra $V(k)$, the regression matrix $K$ consists of regressors $B_p(k)$ centered around the harmonics $P\mathbb{H}_\mathrm{nl}$, the parameter vector $\theta$ contains the parameters $\theta_q$, $q=0,1,...,N_q$, $\theta_p(k')$, $p=0,1,...,N_p$ and $k'\in P\mathbb{H}_\mathrm{nl}$, and $N_V$ a vector representing the noise. The optimal parameters are estimated in linear least squares sense,
\begin{subequations}
\begin{align}
\hat \theta&=\arg \min_\theta (V-K\theta)^H(V-K\theta)\\
&=(K^HK)^{-1}K^HV.
\end{align}
\end{subequations}
For long data records, the regression problem becomes too large to solve in one go. Therefore, it is proposed to solve it in local frequency bands as detailed in \cite{hallemanstimevarying}. We retrieve the impedance and nonlinear distortion estimates from the estimated parameter vector $\hat \theta$,
\begin{subequations}
\begin{align}
\hat Z_p(\omega_k)&=\frac{\hat \theta_p(k)}{I_\mathrm{exc}(k)} & &k\in P\mathbb{H}_\mathrm{exc}\\
\hat V_{\mathrm{s},p}(k)&=\hat \theta_p(k) & &k\in P\big(\mathbb{H}_\mathrm{nl}\setminus\mathbb{H}_\mathrm{exc}\big)
\end{align}
\end{subequations}
Finally, the estimate of the time-varying impedance at the excited frequencies is obtained as,
\begin{align}
\hat Z(\omega_k,t)=\sum_{p=0}^{N_p}\hat Z_p(\omega_k)b_p(t)& & k\in P\mathbb{H}_\mathrm{exc}.
\end{align}
Odd nonlinear distortions and noise introduce uncertainties on the time-varying impedance estimates $\hat Z(\omega_k,t)$. For the computation of uncertainty bounds due to nonlinear distortions and noise, the reader is referred to \cite{hallemanstimevarying}, where also a noise estimation is performed. 

The strength of operando EIS is that it computes the BLTVA of NLTV data, together with uncertainty bounds. Hence, electrochemical systems not satisfying linearity can still be monitored using an impedance. However, the odd nonlinearities must not be too strong. Moreover, since the drift signal is modelled as well, low frequency information becomes available, which is important for some applications. Therefore, it is applicable to a wide range of experiments. Also, it does measure the actual definition of the time-varying impedance, or the BLTVA in the nonlinear case.

The trade-off between time- and frequency resolution, however, remains. To be able to extract the time-variation, we should leave enough empty DFT lines in between the excited lines (corresponding to measuring a large integer number of periods), which decreases the resolution of the excited lines.

\section{A case study on commercial Li-ion batteries}\label{Section:caseStudyLiIon}
Li-ion batteries are chosen as a case study to illustrate the important concepts in this article on real-life measurements. Similar experiments are performed as in \cite{hallemans2022operando}. Measurements are performed on a pristine Samsung INR21700-48X cell, placed in a thermal chamber at $5^\circ$C or $25^\circ$C. The commercially available Samsung 48X is a \SI{4.8}{Ah} $21$ $700$ cell format with cathodes based on lithiated metal oxide (Co, Ni, Al) and anodes based on intercalation graphite and blended Si.

Current and voltage data are collected using the Gamry Interface 5000E potentiostat. Besides running classical EIS experiments, this potentiostat allows to apply user-defined excitations, and measure current and voltage data. The sampling frequency for these user-defined excitations is limited to \SI{200}{Hz}. The current range that can be applied is limited to $[-5,5]$~A.

An odd random phase multisine signal $i_\mathrm{exc}(t)$ is designed with period $T_p=3$~min. The $76$ excited frequencies are chosen as odd harmonics, $\Hexc=\{1,3,5,...\}$, logarithmically distributed between $f_\mathrm{min}=5.6\,\mathrm{mHz}$ and $f_\mathrm{max}=80\,\mathrm{Hz}$. The phases are chosen randomly, uniformly distributed in $[0,2\pi)$. Since noise is often more prominent at low frequencies, and while doing operando experiments the low frequency content is hidden by drift signals, the amplitudes of the multisine are chosen with a decreasing shape over frequency, with root-mean-square (RMS)\nomenclature[A]{RMS}{Root-mean-square} of $0.8$~A rms. 
\subsection{Estimating classical impedance data}
\begin{figure}
    \centering
    \includegraphics[width=0.5\textwidth]{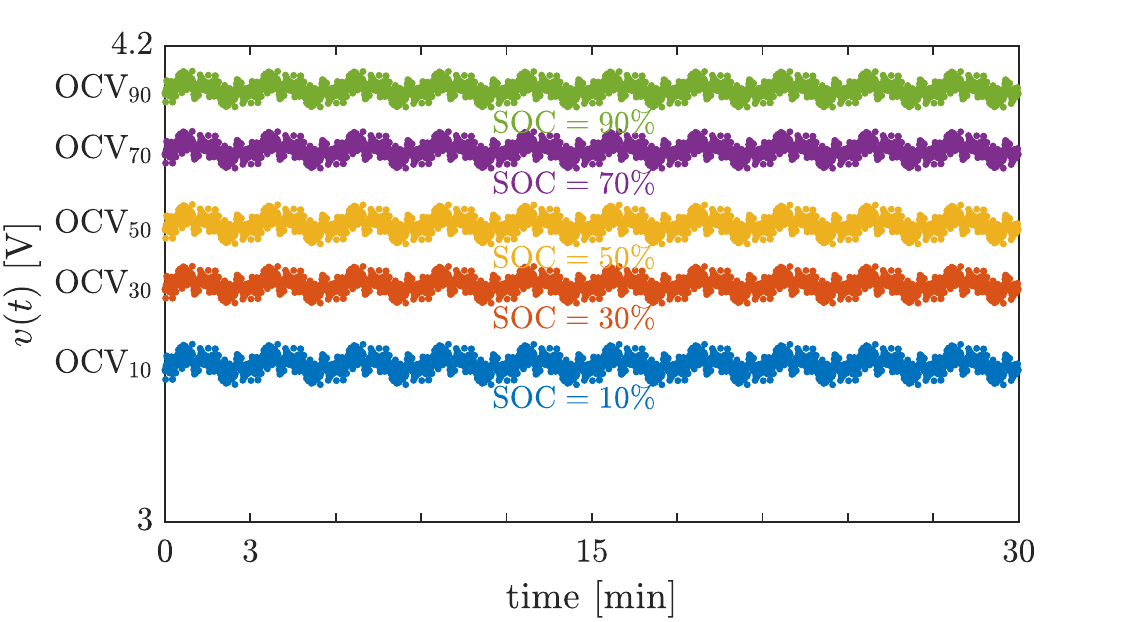}
    \caption{Voltage response of the Samsung 48X cell in LTI conditions at $25^\circ$C. The zero-mean excitation signal $i_\mathrm{exc}(t)$ is used and $P=10$ periods of the current and voltage are measured at different SOC levels. Every color indicates a measurement at a different SOC. Note that for every SOC, the OCV is also different. The time-series are subsampled $150$ times.}
    \label{fig:voltageResponseDifferentOCVLiion}
\end{figure}

For \emph{classical} EIS experiments, the battery is first entirely discharged, then charged at a $C/3$ rate (constant current of $4.8/3=1.6$~A) until the desired SOC level ($10,20, \hdots,90$\%). Two hours of relaxation are allowed such that the battery reaches steady-state and the voltage reaches the OCV value. Then, the zero-mean excitation signal $i(t)=i_\mathrm{exc}(t)$ is looped for $P=10$ periods, that is for $T=PT_p=30$~min. The measured voltage at the different operating points ($\text{SOC}=10,30,50,70,90$\% and $\mathrm{T}=25^\circ$C) is shown in Fig.~\ref{fig:voltageResponseDifferentOCVLiion}. Note that indeed, the voltage data looks periodic with period $3$~min, and that the data is nicely centered around the OCV value, as should be the case in LTI measurements. Since the measurements are periodic, the classical impedances are easily computed from Section~\ref{Section:classicalImpedanceEstimationFD} \eqref{eq:impedanceEstimateLTIPeriodicVoverI}.

The spectra of the measured current and voltage, and estimated impedance for the $10$\% SOC and $25^\circ$C operating point are shown in Fig.~\ref{fig:MeasurementLiionIOspectrumLTI}. Note that for the chosen excitation amplitudes, the battery behaves very linearly. No nonlinear distortions nor nonstationarities can be detected. However, a (small) drift signal is present. Also, noise is clearly present in the measurements, but it is fairly low, an SNR of at least $1000$ is obtained.

The estimated impedances at different operating points (depending on SOC and temperature) are shown in Fig.~\ref{fig:cEISOCV}. We do indeed obtain a different impedance for each of the operating points. Note that if we want to perform quicker experiments, we can measure less periods, for instance $P=4$, leading to experiments of $T=12$~min, however, with a higher noise floor.
\subsection{Estimating time-varying impedance data}
\begin{figure}
    \centering
    \includegraphics[width=0.5\textwidth]{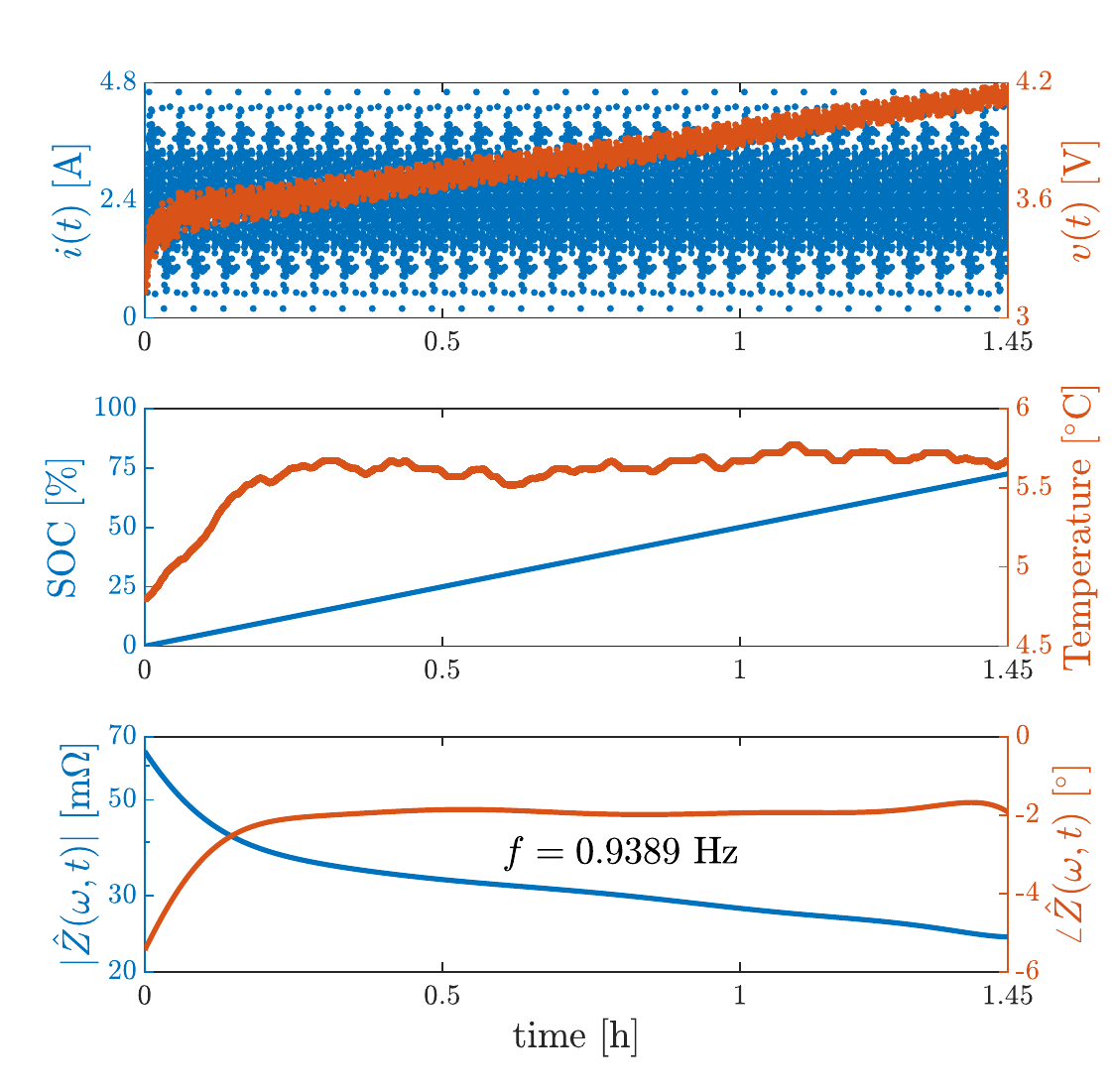}
    \caption{Experiment performed on a Samsung 48X cell in time-varying conditions in a thermal chamber at $5^\circ$C. Top graph: current excitation and voltage response. The current has a DC offset of $2.4$~A, hence, the battery charges, and the voltage increases. Middle graph: SOC, obtained by Coulomb counting, and the external parameter temperature during the experiment. Since the battery is charging with a constant current plus zero-mean multisine, the SOC increases linearly and the temperature increases slightly. Bottom graph: time-variation of the impedance at $0.9389\,\mathrm{Hz}$ obtained from operando EIS \cite{hallemans2022operando}.}
    \label{fig:measurementOperandoTimeDomain}
\end{figure}

For obtaining \emph{time-varying} impedance data, the battery is charged with a $C/2$ current, with the multisine superimposed, 
\begin{align}
i(t)=2.4\,\mathrm{A} + i_\mathrm{exc}(t).
\end{align}
The top graph of Fig.~\ref{fig:measurementOperandoTimeDomain} shows the measured current and voltage data at $5^\circ$C. Due to the DC offset of $2.4$~A in the current, the battery is charging, and the voltage goes up, leading to a drift signal superimposed on the multisine response. The measurement is stopped when the voltage hits the safety bound of $4.2$\,V. For a constant current charging of $C/2$ we would expect the $4.2$\,V to be reached after $2$\,h. However, due to the multisine added on top of this charging current the safety limit is reached prematurely. Accordingly, $P=29$ and $P=31$ periods of the excitation could be measured for the $5^\circ$C and $25^\circ$C experiments, respectively.

The middle graph of Fig.~\ref{fig:measurementOperandoTimeDomain} shows the SOC, with values from $0$~\% to $72.5$~\%, and the battery's surface temperature, which increases slightly due to the charging current. This was also shown in the SOC-temperature plane in Fig.~\ref{fig:operatingTrajectoryImpedance}.

\begin{figure}
    \centering
    \includegraphics[width=0.5\textwidth]{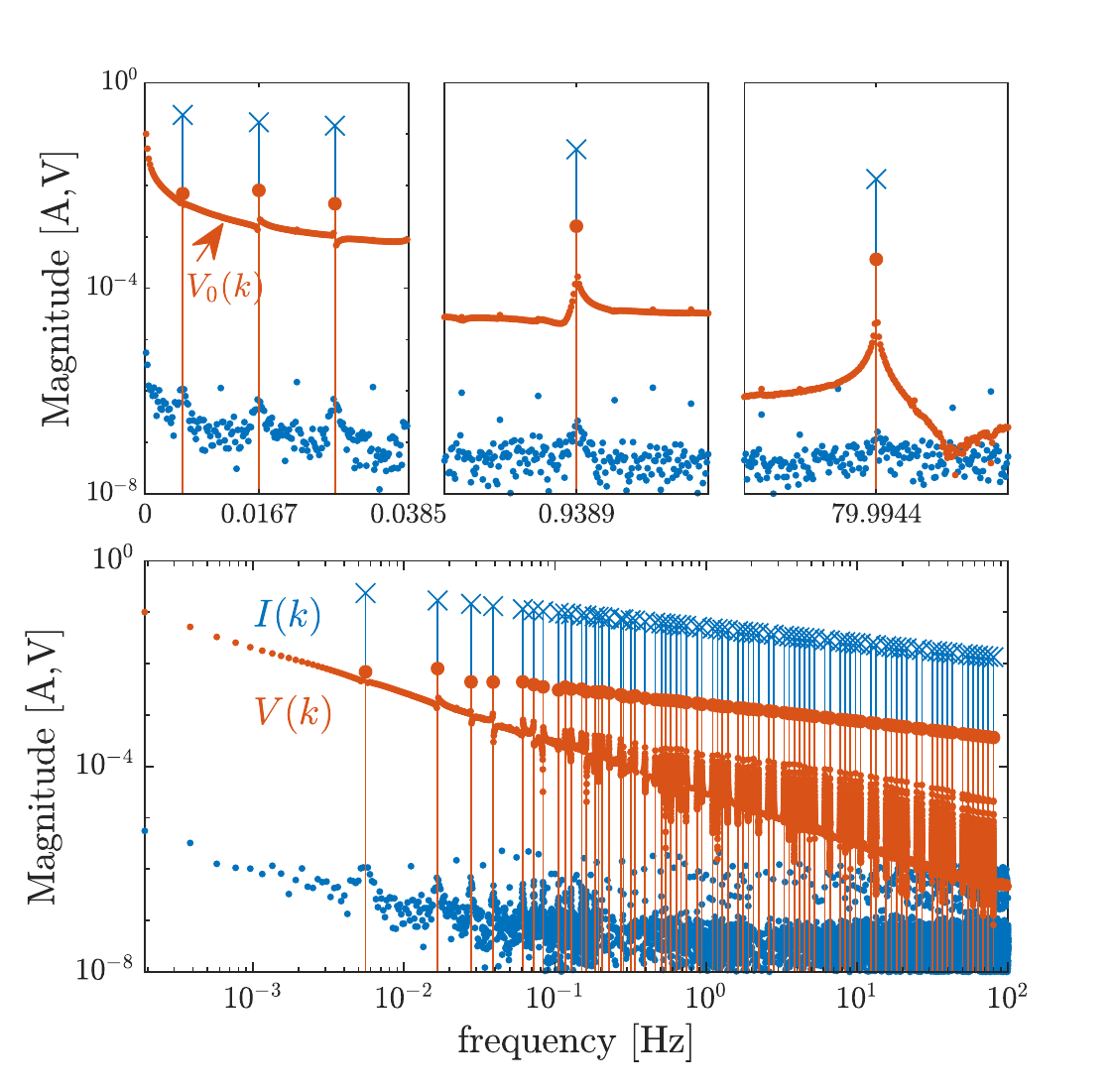}
    \caption{Current and voltage spectra of an experiment performed on a Samsung 48X cell in time-varying conditions at $5^\circ$C. Top graph: spectra of current and voltage in three different frequency bands of each $38.5\,\mathrm{mHz}$ wide, with a linear frequency axis. Note the decreasing shape of the drift spectrum hiding low frequency content. Bottom graph: entire spectra, with a logarithmic frequency axis.}
    \label{fig:MeasurementLiionIOspectrumTimeVarying}
\end{figure}

The spectra of the current and voltage data of the $5^\circ$C experiment are shown in Fig.~\ref{fig:MeasurementLiionIOspectrumTimeVarying}. The bottom graph shows the entire spectra with a logarithmic frequency axis, while the top graph shows zoomed spectra in different frequency bands, each $36\,\mathrm{mHz}$ wide, with a linear frequency axis. Note the general decreasing shape of the drift spectrum $V_0(k)$ in the voltage spectra which hides the low frequency content. For the lowest zoomed frequency band (top left), the time-invariant contributions at the excited frequencies barely exceed the drift spectrum, and the skirts are completely hidden. At frequencies close to $1\,\mathrm{Hz}$, the skirt around the excited frequency is a little more visible, still the drift spectrum hides information. At frequencies close to $80\,\mathrm{Hz}$ the skirts are clearly visible.

The time-varying impedance at $1\,\mathrm{Hz}$, estimated using operando EIS \cite{hallemans2022operando}, is shown in the bottom graph of Fig.~\ref{fig:measurementOperandoTimeDomain}. The time-varying impedance at all excited frequencies is shown in Fig.~\ref{fig:operatingTrajectoryImpedance}. Even though the drift signal hides the low frequency content, clean impedance data can be obtained at these frequencies using operando EIS. Note that the impedance is highest at low SOC, and that the impedance while charging is different from while resting, as also observed in \cite{hallemans2022operando}.
\subsection{Estimating equivalent circuit model parameters}
\begin{figure}
    \centering
    \begin{subfigure}[b]{0.5\textwidth}
    \centering
    Equivalent circuit model
    \includegraphics[width=\linewidth]{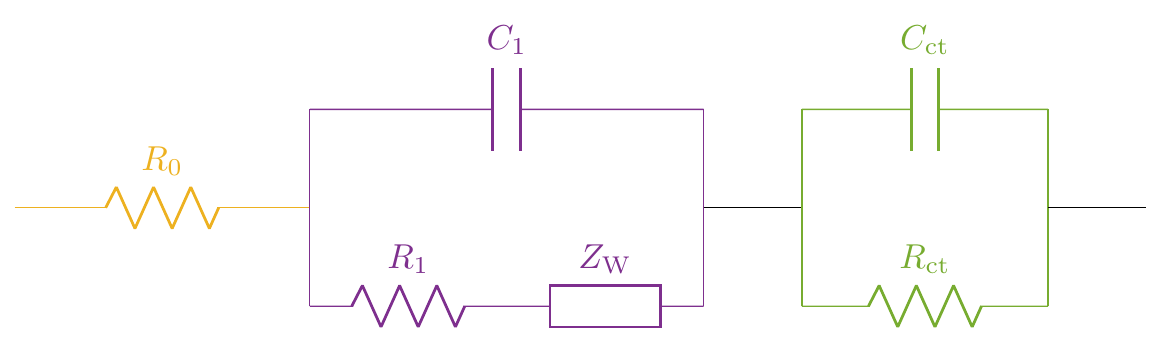}
    \end{subfigure}
    \begin{subfigure}[b]{0.5\textwidth}
    \centering
    Corresponding Nyquist plot and estimated parameters
    \includegraphics[width=\linewidth]{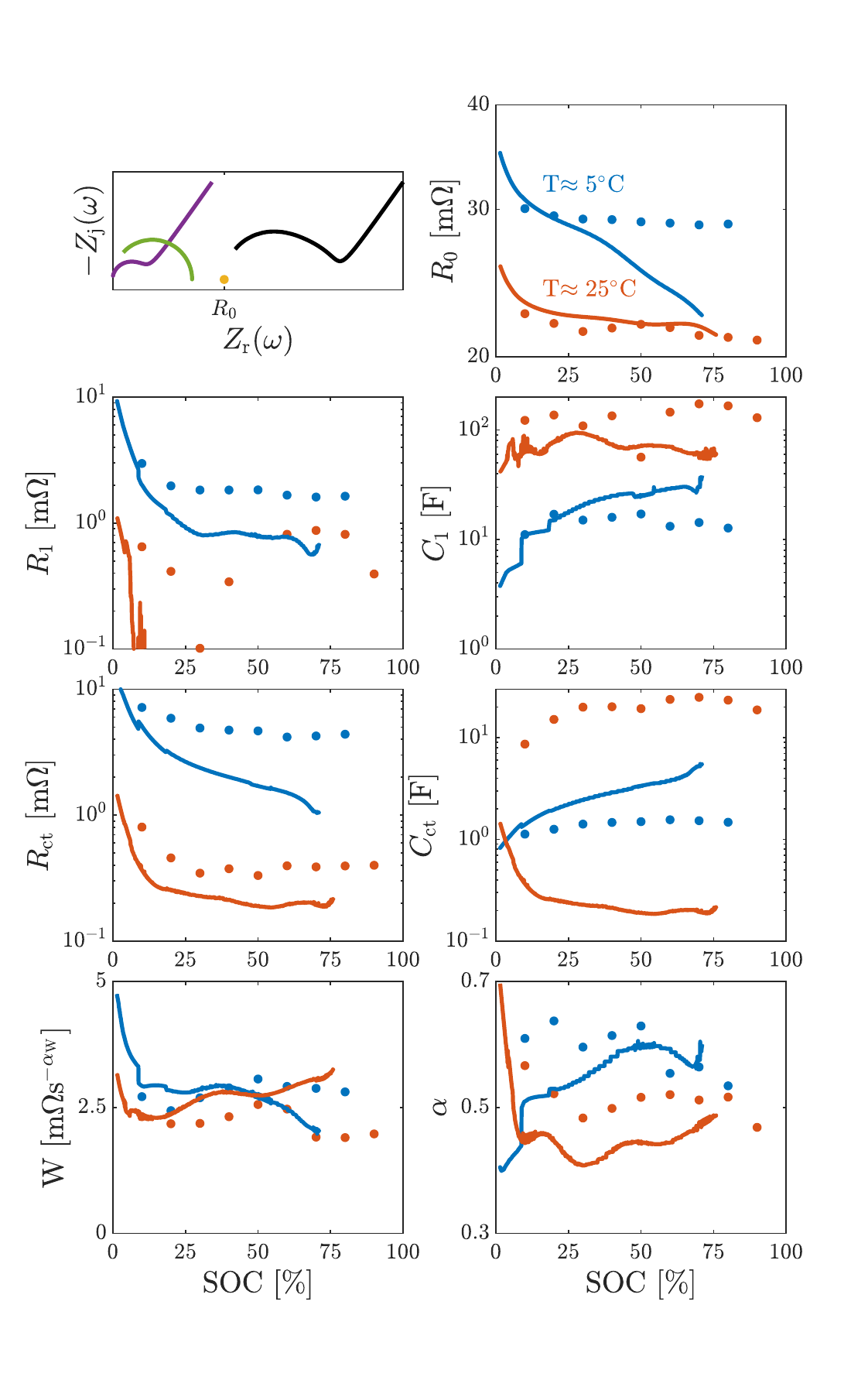}
    \end{subfigure}
    \caption{Equivalent circuit model and estimated parameters for the Samsung 48X cell. The Nyquist plot shows the influence of the different branches (yellow, purple and green) on the total impedance (black). In the ECM parameter graphs, dots represent the parameters obtained from classical EIS at different operating points, while the continuous lines show the parameters obtained from time-varying impedance data along operating trajectories. For the time-varying experiments, the temperature is assumed approximately constant at $5$ or $25^\circ$C.}
    \label{fig:ECMparams}
\end{figure}

Studying the processes going on at different time-scales in the impedance data is often performed by mapping the data onto ECM parameters. For the measured battery impedance data on Samsung 48X cells, we choose the ECM of Fig.~\ref{fig:ECMparams}. This ECM can be linked to the SPM for batteries \cite{kirk2022nonlinear}. The resistance $R_0$ (yellow) is related to the electrolyte resistance, the first $RC$-branch with Warburg element (purple) is related to the diffusion and the second $RC$-branch (green) to the electrochemical kinetics. The corresponding parametric impedance yields,
\begin{subequations}
\begin{align}
    Z_\mathrm{ECM}(\omega,\theta)=R_0+Z_{C_1}\text{//}(R_1+Z_\mathrm{W})+Z_{C_\mathrm{ct}}\text{//}R_\mathrm{ct},
\end{align}
where the parameter vector $\theta$ is given by
\begin{align}
\theta=[R_0,R_1,C_1,R_\mathrm{ct},C_\mathrm{ct},\mathrm{W},\alpha],
\end{align}
the symbol `//' stands for the parallel connection, that is,
\begin{align}
Z_X(\omega)\text{//}Z_Y(\omega)=\frac{Z_X(\omega)Z_Y(\omega)}{Z_X(\omega)+Z_Y(\omega)},
\end{align}
and the impedance of a capacitor and Warburg element, respectively, yield,
\begin{align}
    Z_C(\omega)=\frac{1}{Cj\omega} \quad \text{and} \quad Z_\mathrm{W}(\omega)=\frac{\mathrm{W}}{(j\omega)^\alpha}.
\end{align}
\end{subequations}
The Nyquist chart in Fig.~\ref{fig:ECMparams} illustrates the contribution of the three branches in series (yellow, purple and green) on the total impedance (black), being the sum of the three other colors.

The ECM parameters $\theta$ can now be estimated from impedance data by minimising the cost function,
\begin{align}
\hat \theta = \arg \min_\theta \sum_{k\in P\Hexc}\vert \hat Z(\omega_k)-Z_\mathrm{ECM}(\omega_k,\theta)\vert^2.
\end{align}
This cost function is nonlinear in the parameters $\theta$, hence, a nonlinear solver is required. Here, we use a hybrid of \emph{particle swarm optimisation} \cite{kennedy1995particle,poli2007particle,particleSwarmOptimisation} and the built-in MATLAB function \texttt{lsqnonlin}. Fits over the frequency band [16.7 mHz,50 Hz] are obtained with mean relative errors over frequency all smaller than $0.2$\% and $0.5$\% for the classical and time-varying impedance data, respectively. 

The estimated ECM parameters for the measured impedance data of the Samsung 48X cell at $5^\circ$C and $25^\circ$C are shown in Fig.~\ref{fig:ECMparams}. The ECM parameters of the classical impedance data at different operating points are shown as dots, while the ECM parameters of the time-varying impedance data along trajectories are shown as continuous lines. The temperature is assumed approximately constant during the experiments. It is observed that the parameters obtained in operating and classical conditions are not necessarily equal. For Li-ion batteries, this was already studied in \cite{huang2015dynamic}, where the charge transfer resistance $R_\mathrm{ct}$ at a certain SOC is smaller while charging than while resting.

As an application, Zhu et al.\ \cite{zhu2022operando} propose a fast charging protocol by applying a charging current inversely proportional to the time-varying charge-transfer resistance $R_\mathrm{ct}$, tracked using operando EIS.

\section{Conclusions \& outlook}\label{Section:conclusions}
Classical EIS provides impedance data of electrochemical systems at selected frequencies. Due to the constraints of linearity and stationarity, the impedance data is only valid for small amplitude excitations and at fixed operating points. Nonetheless, measuring classical impedance data is a powerful tool for monitoring electrochemical systems.

Models beyond linearity and stationarity, such as nonlinear leading-order and time-varying impedance, reveal higher-dimensional impedance data, valid over larger excitation amplitudes and along operating trajectories. This higher-dimensional impedance data contains additional information to classical impedance data, which is promising for electrochemical applications. One could, for instance, increase the accuracy of health forecasting of Li-ion batteries using nonlinear and/or time-varying impedance data as indicator. This is also extendable to other electrochemical applications, such as detecting corrosion or studying coatings.

It is shown that the multisine excitation is a strong asset for modelling electrochemical systems. It allows  nonlinear and nonstationary behaviour to be detected from the measured current and voltage data. If this current and voltage data does not satisfy the linearity and stationarity constraints, higher-dimensional impedance data can still be extracted.

\nomenclature[V]{$k$}{DFT index}

\printnomenclature
\section*{Acknowledgements}
This article is dedicated to Rik Pintelon. NH has received funding from the Eutopia mobility programme. NH and JL are supported financially by the Fund for Scientific Research (FWO Vlaanderen) and the Flemish government, Belgium (grant number: METH1). FLM has received funding from the European Research Council (ERC) under the European Union's Horizon 2020 research and innovation programme (grant agreement n° 772579). This work was supported by the Fraunhofer Internal Programs under Grant No. Attract 028-602604 ProLIBs.

\appendix

\section{The Fourier transform}
\label{app:FT}
\begin{align}
&X(\omega)& &=\mathcal{F}\{x(t)\}& &=\int_{-\infty}^{\infty}x(t)e^{-j\omega t}\mathrm{d}t\\
&x(t)& &=\mathcal{F}^{-1}\{X(\omega)\}& &=\int_{-\infty}^{\infty}X(\omega)e^{j\omega t}\mathrm{d}\omega.
\end{align}

\section{Volterra series coefficients}
\label{app:VolterraSeriesCoefficients}
Second order nonlinearity
\begin{align}
V_{2,0}&=\frac{1}{4}\big(Z_2(-\omega,\omega)+Z_2(\omega,-\omega)\big)I^2\\
V_{2,1}&=0\\
V_{2,2}&=\frac{1}{2}Z_2(\omega,\omega)I^2
\end{align}
Third order nonlinearity
\begin{align}
V_{3,0}&=0\\
V_{3,1}&=\frac{1}{8}\big(Z_3(-\omega,\omega,\omega)+Z_3(\omega,-\omega,\omega)+Z_3(\omega,\omega,-\omega)\big)I^3\\
V_{3,2}&=0\\
V_{3,3}&=\frac{1}{8}Z_3(\omega,\omega,\omega)I^3
\end{align}
\section{Linearising an NLTI system around an operating trajectory}
\label{app:linearisingOperatingTrajectory}
The origin of the nonstationarity could be proven by applying a particular excitation to a Volterra series, consisting of a slow part, dictating the trajectory, and a fast part, which is the excitation, 
\begin{align}
i(t)=\underbrace{i_0(t)}_\text{slow}+\underbrace{i_\text{exc}(t)}_\text{fast}.
\end{align}
As an example, the slow part could be a positive constant current for charging a battery, and the fast part a multisine. By assuming that the fast perturbation has a small amplitude, and hence, only the linear part of the Volterra series ($n=1$) is needed with respect to $i_\text{exc}(t)$, the voltage response $v(t)$ can also be separated into a slow and fast part,
\begin{subequations}
\begin{align}
v(t)=v_0(t)+v_\text{exc}(t),
\end{align}
with
\small
\begin{align}
v_0(t)& = \mathrm{OCV}+\sum_{n=1}^{n_\mathrm{max}}\int_{-\infty}^\infty \cdots \int_{-\infty}^\infty z_n(\tau_1,...,\tau_n)\prod_{l=1}^n i_0(t-\tau_l)\mathrm{d}\tau_l \nonumber \\
v_\text{exc}(t)& = \int_{-\infty}^\infty \underbrace{z(\tau,t)}_{\text{depends on }z_n \text{'s and }i_0(t)} i_\text{exc}(\tau)\mathrm{d}\tau.
\label{eq:separationdcacVoltage}
\end{align}
\normalsize
\end{subequations}
The slow part $v_0(t)$\nomenclature[F]{$v_0(t)$}{Drift signal} is called the drift signal, and solely depends on the slow excitation $i_0(t)$. The fast response is now the convolution of a two-dimensional impulse response $z(\tau,t)$ with the excitation. This two-dimensional impulse response function explicitly depends on the time of excitation $t$, such that stationarity is not satisfied anymore. Moreover, this function is shown to depend on the generalised impulse responses $z_n(\tau_1,\dots,\tau_n)$ and the slow signal $i_0(t)$.

\section{The discrete Fourier transform}
\label{app:DFT}
\begin{align}
& X(k)  & &= \frac{1}{N}\sum_{n=0}^{N-1}x(nT_s)e^{-j\frac{2\pi k n}{N}}\\
& x(nT_s) & &= \sum_{k=0}^{N-1}X(k)e^{j\frac{2\pi k n}{N}}
\end{align}

\section{Equivalence between \eqref{eq:impedanceDarowicki} and \eqref{eq:impedanceLaMantia}}
\label{app:Equivalence}
Using the following properties of the Fourier transform,
\begin{subequations}
\begin{align}
X(\omega)=\mathcal{F}\{x(t)\} \qquad x=x,y,w,i,v
\end{align}
\begin{align}
\mathcal{F}\{x(t)y(t)\}=X(\omega)\ast Y(\omega)=\int_{-\infty}^\infty X(\omega-\omega')Y(\omega')\mathrm{d}\omega'
\end{align}
\begin{align}
\mathcal{F}\{x(t'-t)\}=X(\omega)e^{-j\omega t},
\end{align}
\end{subequations}
where $\mathcal{F}$ acts on $t'$, one finds that,
\begin{align}
\mathcal{F}\{w(t'-t)x(t')\}&=\int_{-\infty}^\infty W(\omega-\omega')e^{-j(\omega-\omega')t}X(\omega')\mathrm{d}\omega'\nonumber \\
&=e^{-j\omega t} \int_{-\infty}^\infty W(\omega-\omega')X(\omega')e^{j\omega' t}\mathrm{d}\omega'\nonumber \\
	&= e^{-j\omega t} \mathcal{F}^{-1}\{W(\omega-\omega')X(\omega')\}.
\end{align}
Assuming that $w(t)=w(-t)$, one has that $W(\omega)=W(-\omega)$, accordingly,
\begin{align}
\frac{\mathcal{F}\{w(t'-t)v(t')\}}{\mathcal{F}\{w(t'-t)i(t')\}}&=\frac{\mathcal{F}^{-1}\{W(\omega-\omega')V(\omega')\}}{\mathcal{F}^{-1}\{W(\omega-\omega')I(\omega')\}}\nonumber \\
& =\frac{\mathcal{F}^{-1}\{W(\omega'-\omega)V(\omega')\}}{\mathcal{F}^{-1}\{W(\omega'-\omega)I(\omega')\}}
\end{align}
with all Fourier and inverse Fourier transforms acting on, respectively, $t'$ and $\omega'$.
\section{Legendre polynomials}
\label{app:LegPol}
The Legendre polynomials $L_p(x)$, $p=0,1,...$, $x\in [-1,1]$ are the solution of Legendre's differential equation
\begin{align}
\frac{\mathrm{d}}{\mathrm{d}x}\Big((1-x^2)\frac{\mathrm{d}L_p(x)}{\mathrm{d}x}\Big)+p(p+1)L_p(x)=0.
\end{align}
The basis functions $b_p(t)$ are chosen as rescaled Legendre polynomials over the interval $[0,T]$, that is,
\begin{align}
b_p(t)=L_p\Big(\frac{2t}{T}-1\Big).
\end{align}

\bibliographystyle{elsarticle-num}
\bibliography{bibFile}

\end{document}